\documentclass[amsmath,amssymb,pra,twocolumn]{revtex4}

\usepackage{graphics}
\usepackage{epsfig}
\usepackage{amsmath}
\usepackage{color}
\usepackage{dcolumn}
\usepackage{bm}
\usepackage{multirow}

\newcommand{\ds }{}

\newcommand{\Li}{$^{6}$Li }

\newcommand{\LiRb}{$^{6}$Li-$^{87}$Rb }
\newcommand{\Rb}{$^{87}$Rb }

\newcommand{\ket}[1]{| #1 \rangle}
\newcommand{\bra}[1]{\langle #1 |}

\def\oper#1{\hat{\rm #1}}

\def\dC6{\delta\hspace{-0.2mm}C_6}
\def\dG{\delta\hspace{-0.2mm}G}
\def\dm{\delta\hspace{-0.2mm}\mu}

\def\eg{e.\,g.}
\def\ie{i.\,e.}
\def\etal{{\it et al.} }

\begin{document}

  \bibliographystyle{apsrev}

  \title{Mimicking multi-channel scattering with single-channel approaches}

       \author{Sergey Grishkevich, Philipp-Immanuel Schneider, 
               Yulian V.~Vanne and Alejandro Saenz}

       \affiliation{AG Moderne Optik, Institut f\"ur Physik,
         Humboldt-Universit\"at zu Berlin, Hausvogteiplatz 5-7,
         10117 Berlin, Germany}

       \date{\today}

  \begin{abstract}
    The collision of two atoms is an intrinsic multi-channel (MC) problem as
    becomes especially obvious in the presence of Feshbach resonances. Due to
    its complexity, however, single-channel (SC) approximations, which
    reproduce the long-range behavior of the open channel, are often applied
    in calculations. In this work the complete MC problem is solved
    numerically for the magnetic Feshbach resonances (MFRs) in collisions
    between generic ultracold ${}^6$Li and ${}^{87}$Rb atoms in the ground
    state and in the presence of a static magnetic field $B$. The
    obtained MC solutions are used to test various existing as well as
    presently developed SC approaches. It was found that many aspects even at
    short internuclear distances are qualitatively well reflected. This can be
    used to investigate molecular processes in the presence of an external
    trap or in many-body systems that can be feasibly treated only
    within the framework of the SC approximation. The applicability of various
    SC approximations is tested for a transition to the absolute vibrational
    ground state around an MFR. The conformance of the SC approaches is
    explained by the two-channel approximation for the MFR.
  \end{abstract}

    \maketitle

 \section{Introduction}
 \label{sec:intro}

 The tunability of the interparticle interaction on the basis of Feshbach
 resonances, especially magnetic ones (MFRs), marked a very important
 corner-stone in the research area of ultracold atomic gases. At ultracold
 energies s-wave scattering dominates the atom-atom interaction, such that for
 large internuclear distances the elastic scattering properties are solely
 described by the s-wave scattering length $a_{\rm sc}$~\cite{cold:wein99}. 
 Its sign determines the type of  interaction (repulsive or attractive) and
 its absolute value the interaction strength. In the presence of an MFR this
 parameter can be tuned at will by applying an external magnetic field. A wide
 range of experiments using MFR techniques has been carried out including the
 formation of cold, even Bose-Einstein condensed
 molecules~\cite{cold:joch03,cold:rega03,cold:zwie03} or the realization of a
 Mott insulator phase with atoms in an optical lattice (OL) \cite{cold:robe08}.
 
 Experiments with ultracold gases are usually performed in external trapping
 potentials and over an ensemble of many particles. For tight trapping
 conditions the influence of the additional potential can become
 essential. For example, processes of molecule formation via photoassociation
 (PA) where two ultracold atoms absorb a photon and form a bound excited
 molecule~\cite{cold:lett93,cold:fior98} can be more efficient, if performed
 under tight trapping conditions as they are accessible in
 OLs~\cite{cold:jaks02,cold:deb03,cold:gris07}.
 
 However, the presence of a trapping potential or, worse, the existence of
 many-body effects is a great challenge for the full theoretical description
 of an MFR, since all accessible spin configurations of the colliding atoms
 must be included, leading to a multi-channel (MC) problem. For the case of
 s-wave scattering of two free atoms the separation in relative and
 center-of-mass motions, the formulation in spherical coordinates, and
 the continuous energy spectrum make the numerical solution manageable. This
 changes, unfortunately, if an external potential couples the six spatial
 coordinates of the two colliding atoms and induces the need to find discrete
 eigenenergies~\cite{cold:gris09}. This can be the case for atoms loaded in a
 cubic OL formed with the aid of standing light
 waves~\cite{cold:jaks98,cold:grei02,cold:koeh05}. Furthermore, the 
 theoretical microscopic investigation of ultracold many-body systems is
 feasible only within the framework of the SC approximation. Nevertheless, a
 good knowledge of two-body MC collisions should help in understanding the
 consequences of SC approximations which must be done when many-body systems
 are considered.

 Single-channel (SC) approximations allowed to study the influence of 
 the scattering length as it results from an MFR for three-body collisions 
 \cite{cold:esry99a,cold:simo09} 
 and in the presence of an external trap \cite{cold:deb03, cold:gris07, cold:schn09}.
 However, to our knowledge it is not yet well established to what extent SC
 approximations describe correctly the behavior of a coupled MC system, if
 more than one channel contributes significantly. The successful usage of (SC)
 pseudo-potentials to model, \eg, the atom-atom interaction in
 OLs~\cite{cold:bloc08} shows that physical properties depending on the
 long-range behavior of the open-channel scattering wave function, \ie, the
 scattering length $a_{\rm sc}$, are well described within the SC
 framework. For shorter interatomic distances in the order of the van der
 Waals length scale $\beta_6$ ($\ds \beta_6 = (2\mu C_6)^{1/4}$ where $\mu$ is
 the reduced mass and $C_6$ is the van der Waals coefficient) this is not
 necessarily the case. Here, all coupled channels contribute to the full wave
 function and affect processes,  such as transitions to molecular bound
 states. For these distances, SC approximations cannot cover all details of
 the MC solution. As will be shown, some important aspects are, nevertheless,
 reflected and can be used to study processes of molecule formation in the
 presence of an MFR where MC calculations may be too laborious. A very
 systematic investigation of both short-range and long-range parts of the MC
 solutions against various SC ones is considered  that is done in this work.

 The formation of ultracold molecules especially in deeply bound levels is
 currently of large interest. In order to associate them, the starting point
 is often a sample of Feshbach molecules, obtained from ultracold atoms via a
 sweep of the magnetic field around an
 MFR~\cite{cold:inou98,cold:rega03,cold:koeh06}. These molecules are usually
 formed in high lying vibrational ground states. Molecules in lower vibrational
 states and eventually in the absolute vibrational ground state are, however,
 favorable since they are more stable against inelastic collisions. The most
 successful scheme to access those molecules is the two-color stimulated Raman
 adiabatic passage (STIRAP)~\cite{cold:drum02} when the passage is realized using
 PA via an intermediate excited state. The dump photoassociation (DPA) process
 during which two ultracold atoms absorb a photon and form directly a ground
 molecule is in principle possible for heteronuclear systems, although the
 yield is very small.

 It has been shown theoretically and for some cases even experimentally, that
 the PA and DPA yields can be significantly increased in the presence of an
 MFR~\cite{cold:abee98,cold:cour98,cold:rega03,cold:gris07,cold:junk08,cold:pell08,cold:deig09}.
 For example in~\cite{cold:gris07} it was found that an SC scheme
 based on mass variation predicts the same enhancement of the PA rate
 for almost all final states except the very high-lying ones and the ones at the
 PA window (for $a_{\rm sc}>0$). The reason for the enhancement was the
 increase of the absolute value for the initial-state wave function that occurs
 for large absolute values of $a_{\rm sc}$. As a consequence the corresponding
 Franck-Condon (FC) factors and PA yields increase with $|a_{\rm sc}|$ (see
 Sec.III,G of~\cite{cold:gris07} for details). Noteworthy, a strong
 enhancement of the PA rate by at least two orders of magnitude
 while scanning over an MFR was predicted on the basis of a MC
 calculation for a specific $^{85}$Rb resonance already
 in~\cite{cold:abee98}. The explanation for the enhancement given
 in~\cite{cold:abee98} is, however, based on an increased admixture of a
 bound-state contribution to the initial continuum state in the vicinity of
 the resonance. This is evidently different from the reason for the
 enhancement due to large values of $|a_{\rm sc}|$ discussed
 in~\cite{cold:gris07}. This suggests that both seemingly different systems
 appear exhibit a strong correspondence. One of the motivations of the present
 work is to clarify this observation.

 To mimic certain aspects of the MC wave function for studying molecular
 processes, SC approaches make use of a controlled tuning of system
 parameters such as the reduced mass~\cite{cold:gris07}, van der Waals
 coefficients~\cite{cold:ribe04}, inner wall~\cite{cold:gris09} of the
 interaction potential or the interaction potential in the intermediate range
 as is proposed in this work. Long-range scattering properties like the s-wave
 scattering length can be sensitive to even small changes of those
 parameters. To date, the justification of these systematical variations is
 mainly given by the broad variety of atomic species and their isotopes, each with
 different parameter values. In this work it will be shown that by these
 variational approaches one is also able to reproduce changes of both long and
 short range collision properties of a given scattering system as it is induced by an
 external magnetic field in the proximity of an MFR. 

 The general validity of the SC methods will be based on a two-channel (TC) approximation of the 
 MFR~\cite{cold:fesh58,cold:koeh06}. This approximation is widely used to describe the phenomenon 
 of an MFR and has been adopted to study many-body interactions \cite{cold:kokk02} and two-atom 
 interaction in a time-dependent magnetic field  \cite{cold:mies00,cold:gora04} and in a structured 
 continuum induced by an OL \cite{cold:nyga08}.
 The TC approximation reproduces many aspects of the coupled MC system. It allows to 
 describe the complex PA transition process by just two free parameters, the maximal 
 transition rate and the position of the minimal transition rate \cite{cold:schn09a}. 
 An analysis of the TC approximation reveals why SC approaches can show an
 astonishing conformance with the coupled MC predictions.

 In order to compare concrete MC and SC solutions the exemplary case of 
 \Li and \Rb scattering is considered and the relative motion of this system in a
 static magnetic field $B$ is fully solved employing the $R$-matrix
 method~\cite{cold:burk07}. This system is of great importance by itself for
 its large static dipole moment, which makes it interesting for applications
 in quantum information processing~\cite{cold:mich06,cold:rabl06} or the
 exploration of lattices of dipolar molecules~\cite{cold:pupi08}. The
 applicability of the different SC approaches is studied by considering the
 process of molecule formation by a direct PA of \LiRb to the absolute
 vibrational ground state in the presence of an MFR. We describe this process
 by using the exact MC solution and compare to different SC approximations.

 The paper is organized in the following way. In Sec.~\ref{sec:multichappr}, a
 theoretical description of \Li and \Rb scattering is given and the TC
 approximation is briefly introduced. The possibility of SC approaches is
 motivated by considering the results of a full MC calculation for different
 resonant and off-resonant magnetic field values. In Sec.~\ref{sec:SC_appr},
 diverse SC approaches are introduced and their wave functions are compared to
 those of the full MC calculation. The direct dumping to the absolute
 vibrational ground state is considered in Sec.~\ref{sec:PAtoGS}. The
 prediction of the TC approximation is presented and MC and SC results are
 compared. Finally, a conclusion is given in Sec.~\ref{sec:conclusion}. All
 equations in this paper are given in atomic units unless otherwise specified.

 \section{Multi-channel approach}
 \label{sec:multichappr}

 \subsection{Hamiltonian}
 \label{ssec:Hamiltonian}

 The Hamiltonian of relative motion for two colliding ground-state alkali atoms -- 
 in the present case $^6$Li (atom 1) and $^{87}$Rb (atom 2) -- is given as
 \cite{cold:moer95} 
 \begin{equation}
   \ds
   \oper{H} = \oper{T}_{\mu}+\sum_{j=1}^2
   (\oper{V}_j^{\rm hf}+\oper{V}_j^{\rm Z})+\oper{V}_{\rm int}
   \label{eq:BfieldHam}
 \end{equation}
 where $\oper{T}_{\mu}$ is the kinetic energy and $\mu$
 is the reduced mass. The hyperfine operator $\ds \oper{V}_j^{\rm hf}=
 a_{\rm hf}^j\vec s_j\cdot \vec i_j$ and the Zeeman operator 
 \mbox{$\oper{V}_j^{\rm Z}=(\gamma_e \vec s_j - \gamma_n \vec i_j)\cdot \vec B$} 
 in the presence of a magnetic field $\vec B$ 
 depend on the electronic spin $\vec s_j$ and nuclear spin $\vec i_j$ of atom $j=1,2$. 
 For the present system the values of the hyperfine constants 
 $a_{\rm hf}^1$, $a_{\rm hf}^2$, and those of the nuclear and electronic
 gyromagnetic factors $\gamma_n$ and $\gamma_e$ are adopted 
 from~\cite{cold:arim77}. In Eq.~(\ref{eq:BfieldHam}) the central interaction
 $\oper{V}_{\rm int}(R)$ between the atoms is a combination of electronic singlet
 and triplet contributions
\begin{equation}
  \ds
  \oper{V}_{\rm int}(R) = V_0(R) \oper{P}_0 + V_1(R) \oper{P}_1
  \label{eq:v01}
\end{equation}
 where $\oper{P}_{0}$ and $\oper{P}_{1}$ project on the singlet
 and triplet components of the scattering wave function, respectively. The
 potential curve $V_0$ ($V_1$) for the singlet (triplet) states of \LiRb in Born-Oppenheimer
 (BO) approximation were obtained using data from~\cite{cold:marz09,vdw:li08}
 and references therein. In~\cite{cold:marz09} refined potential parameters
 such as the van der Waals and exchange coefficients, which we use in the
 following, were determined by a comparison of MC calculations
 with experimentally observed resonances. It is important to note that the
 MC approach considered in the present work is formulated in
 relative motion coordinates. This is based on the assumption that the center-of-mass and relative motion of two
 atoms may be decoupled and effects due to coupling may be
 neglected. Furthermore, calculations of the present work assume the BO
 approximation to be valid~\cite{gen:bran03}.

 For the interactions present in Hamiltonian~(\ref{eq:BfieldHam}) the
 projection $M_F$ of the total spin angular momentum $\vec{F}=\vec{f}_1+\vec{f}_2$ 
 on the magnetic field axis  is conserved during the collision. Here,
 $\vec{f}_j=\vec s_j + \vec i_j$  is the total spin of atom $j$. For a given
 $M_F$ of the colliding atoms only spin-states with the same total projection
 of the angular momentum can be excited during the collision. If
 $\{\ket{\alpha}\}_\alpha$ is a complete basis of the electron and nuclear
 spins of the $M_F$-subspace, one may use the function
 \begin{equation}
   \ds \Psi(R) =
    \sum_{\alpha}
    \frac{\psi_{\alpha}(R)}{R}
    \ket{\alpha}
    \label{eq:mcwfansatz}
 \end{equation}
 in order to find the s-wave scattering solution of the stationary Schr\"odinger equation 
 with Hamiltonian~(\ref{eq:BfieldHam}). 
 This {\it ansatz} yields a system of coupled second-order differential equations
\begin{equation}
\label{eq:multyset}
\begin{split}
   \ds \left(
     -\frac{1}{2\mu}\frac{\partial^2}{\partial R^2} + V_{\alpha}(R) 
     + E_{\alpha}(B)-E 
    \right) &\psi_{\alpha}(R) 
\\
   + \sum_{\alpha'} W_{\alpha'\alpha}(R)\, &\psi_{\alpha'}(R) = 0
\end{split}
\end{equation}
 where the channel threshold energies $E_\alpha$,
 the channel potentials $V_\alpha(R)$, and 
 the coupling potentials $W_{\alpha'\alpha}(R)$ 
 depend on the chosen spin basis and will be specified below.
 
 Depending on the spin basis, the scaled channel functions $\psi_\alpha(R)$
 will be used in the analysis instead of the full channel functions
 $\psi_\alpha(R)/R$, while the name ``channel function'' is kept for
 convenience.

 \subsubsection{Atomic basis}

 If the two atoms are far apart from each other, the central interaction
 $\oper{V}_{\rm int}(R)$ may be neglected and the two-body system is described by
 the spin eigenstates $\ket{f_j,m_{f_j}}$ of each atom. In this atomic basis
 (AB) the collision channels $\ket{\alpha}$ are written as a direct product of
 the atomic states $\ds \ket{\chi}=\ket{f_1,m_{f_1}}\ket{f_2,m_{f_2}}$. In
 this case the threshold energy $E_{\chi}(B)$ of channel $\ket{\chi}$ is
 given as the sum of Zeeman and hyperfine energies of the two atoms. The
 channel potential $V_\chi(R)$ in the AB is identical for all channels,
 \begin{equation}
  \ds V_\chi(R) = V_{+}(R) =\frac{V_0(R)+V_1(R)}{2}\,.
 \end{equation}
 The long-range asymptote of $V_{+}$ is described by an attractive van der
 Waals interaction, that in the present case of ${}^6$Li and ${}^{87}$Rb atoms
 in their ground states is given as 
 \begin{equation}
   \ds
   V_{\rm vdW}(R) = -\sum_{n=3}^5\frac{C_{2n}}{R^{2n}}\quad , 
   \label{eq:vdW}
 \end{equation}
 with $C_6=2543$\,a.u., $C_8=228250$\,a.u., and $C_{10}=25\,645\,000$\,a.u.
 The coupling between the channels in the AB is given as  $W_{\chi'\chi}(R)= 
 \bra{\chi'}\oper P_0 -\oper P_1\ket{\chi} V_{-}(R)$ where 
 \begin{equation}
 \ds V_{-}(R) =\frac{V_0(R)-V_1(R)}{2} = \frac12 V_{\rm ex}(R)\,.
 \end{equation}
 The exchange interaction $V_{\rm ex}$ is in the long-range regime very well 
 represented in the Smirnov and Chibisov form~\cite{vdw:smir65}
\begin{equation}
   \ds 
   V_{\rm ex}(R;J_0,\alpha) = J_0\, R^{\frac{7}{\alpha}-1}
                                      e^{-\alpha R}\,.
   \label{eq:exchange}
 \end{equation}
 In Eq.~(\ref{eq:exchange}) $J_0=0.0125$ is a normalization constant and 
 $\alpha=1.184$ depends on the ionization energies
 of each atom. For a given magnetic field $B$ the channel threshold energies
 $E_{\chi}$ and coupling matrix $W_{\chi \chi'}$ are fixed and $V_{-}(R)$
 describes how strongly the different channels $\ket{\chi}$ are coupled.
 
 The total energy $E$ avaliable to the system is the kinetic energy, \ie, the
 energy at a time prior to the interaction when particles are far apart from
 each other. Since the coupling vanishes exponentially, the channels in the AB
 are asymptotically uncoupled. If the threshold energy of a channel either
 lies above or equals the total energy avaliable to the system, 
 $E_\chi(B) \geq E$, the channel is considered to be ``open'', otherwise it is
 ``closed''. Without loss of generality we consider in the following an
 elastic collision where only the channel $\ket{a_1} = \ket{1/2,1/2}\ket{1,1}$
 with the lowest threshold energy is open. The threshold energy $E_{a_1}$
 marks the zero point of the energy scale throughout the paper.

 \subsubsection{Molecular basis}
 \label{subsubsec:molecbas}

 Another possible choice of the spin basis of the channels $\ket{\alpha}$ is the molecular 
 basis (MB) $\ds \ket{\xi}=\ket{S,M_S}\ket{m_{i_1},m_{i_2}}$ where $S$ and
 $M_S$ are the quantum numbers of the total electronic spin and its 
 projection along the magnetic field. Furthermore, $m_{i_1}$ and $m_{i_2}$ are
 the nuclear spin projections of the individual atoms. 
 In the MB the threshold energy $E_\xi(B)$ is equal to the Zeeman energy of the two atoms.
 Depending on the value of $S$, the channel potentials correspond
 to the singlet ($S=0$) or triplet ($S=1$) potential, \ie,  $V_\xi(R)=V_S(R)$.
 While in the AB the coupling $W_{\chi'\chi}$ is strong for small internuclear distances, 
 in the MB the channels are only coupled by the relatively weak hyperfine interaction.
 The coupling $W_{\xi'\xi}=\bra{\xi'}\oper V^{\rm hf}_1 + \oper V^{\rm hf}_2 \ket{\xi}$
 is, on the other hand, present for all internuclear distances, which makes it impossible
 to define open and closed channels in the MB.

 Depending on the distance between the two particles the set of interacting states
 is preferably considered in either of the two bases~\cite{cold:bhat04,cold:bamb02}. 
 The AB of asymptotically uncoupled states is convenient for the description
 of the long-range part of the wave function. The MB is suitable for the
 short-range part where the exchange interaction leads to a strong coupling in
 the AB. While inappropriate for large distances, the MB is the natural choice to
 study molecular processes, such as the association of molecules, which take
 place when the atoms are close to each other. Presently for \LiRb the
 transition from the description in the AB to the MB is appropriate at a
 distance $R_{\rm sh}\approx 20\,a_0$  ($a_0$ is the
 Bohr radius) where the exchange interaction is equal
 to the hyperfine interaction, \ie, where 
 $\Delta E_{\rm hf}({^6\rm Li}) + \Delta E_{\rm hf}({^{87}\rm Rb}) = J_0 R^{\frac7\alpha -1}e^{-\alpha R}$, 
 with $\Delta E_{\rm hf}({^6\rm Li})=228.2$\,MHz and $\Delta E_{\rm hf}({^{87}\rm Rb})=6834.7$\,MHz
 being the hyperfine splittings~\cite{cold:arim77}.

\begin{table}
  \caption{Atomic and molecular basis states 
    of the \LiRb system for the manifold of states with $\ds M_F=3/2$.}
  \begin{ruledtabular}
     \begin{tabular}{clcl}
         index $\ket{\chi}$ & atomic basis & index $\ket{\xi}$ & molecular basis  \\
         \hline
          $\ket{a_1}$ & $\ket{1/2,1/2}\ket{1,1}$  & $\ket{S_1}$ & $\ket{0,0}\ket{1,1/2}$ \\
          $\ket{a_2}$ & $\ket{3/2,1/2}\ket{1,1}$  & $\ket{S_2}$ & $\ket{0,0}\ket{0,3/2}$ \\
          $\ket{a_3}$ & $\ket{3/2,3/2}\ket{1,0}$  & $\ket{T_1}$ & $\ket{1,-1}\ket{1,3/2}$ \\
          $\ket{a_4}$ & $\ket{1/2,1/2}\ket{2,1}$  & $\ket{T_2}$ & $\ket{1,0}\ket{0,3/2}$ \\
          $\ket{a_5}$ & $\ket{1/2,-1/2}\ket{2,2}$ & $\ket{T_3}$ & $\ket{1,0}\ket{1,1/2}$ \\
          $\ket{a_6}$ & $\ket{3/2,3/2}\ket{2,0}$  & $\ket{T_4}$ & $\ket{1,1}\ket{-1,3/2}$ \\
          $\ket{a_7}$ & $\ket{3/2,1/2}\ket{2,1}$  & $\ket{T_5}$ & $\ket{1,1}\ket{0,1/2}$ \\
          $\ket{a_8}$ & $\ket{3/2,-1/2}\ket{2,2}$ & $\ket{T_6}$ & $\ket{1,1}\ket{1,-1/2}$ \\
     \end{tabular}
  \end{ruledtabular}
  \label{tab:atommolbasnote}
\end{table}

 \subsection{Computational details}

 Since for the present case of \LiRb the channel with the lowest threshold
 energy $\ket{a_1} = \ket{1/2,1/2}\ket{1,1}$ is considered as the open
 entrance channel, only channels with the total angular momentum $\ds M_F=3/2$
 are coupled during the collision. All eight coupled atomic and molecular
 basis states are given in Tab.~\ref{tab:atommolbasnote}. 

 The system of eight coupled equations is numerically solved in the AB employing the 
 $R$-matrix method~\cite{cold:burk07}. This method is a general 
 {\it ab initio} approach to a wide class of atomic and molecular collision
 problems. The essential idea is to divide the physical space into two or
 possibly more regions. In each region the stationary Schr\"odinger equation
 may be solved using techniques designed to be optimal to describe the
 important physical properties of that region. The solutions and their
 derivatives are then matched at the boundaries. The transition from
 AB to MB is carried out by a unitary basis transformation.

 The wave function $\Psi$ in Eq.~(\ref{eq:mcwfansatz}) must obey appropriate boundary
 conditions in order to reduce the number of the independent solutions of
 the set of equations in~(\ref{eq:multyset}) to one. The condition 
 $\psi_{\alpha}(0)=0$ ensures that the full wave function does not diverge at $R=0$. 
 Another demand is that functions of the closed channels $\psi_{\chi}(R)$ must
 vanish at $\ds R\rightarrow\infty$. The implementation of these boundary
 conditions allows to solve Eqs.~(\ref{eq:multyset}) leaving one free
 parameter in the solution, \eg, the normalization of the open channel. 
 We chose to scale the open channel function to the $\sin$-normalized
 form
 \begin{equation}
   \ds
   \psi_{a_1}(R)|_{R\rightarrow\infty} = 
   \sin (k\cdot R+ \delta)\,,
   \label{eq:norm}
 \end{equation}
 with $\ds k=\sqrt{2\mu E}$. The phase shift $\delta$ is a result of the
 interaction and is connected via
 \begin{equation}
   \label{eq:delta_vs_scatlen}
   \ds
   \tan(\delta)=-k\cdot a_{\rm sc}\,,
 \end{equation}
 to the s-wave scattering length $\ds a_{\rm sc}$. In order to normalize the
 incoming channel function its asymptotic form is matched using Eq.~(\ref{eq:norm}). The value of $a_{\rm sc}$ is automatically
 determined by the matching procedure. As will become evident in
 Sec.\ref{ssec:Two-channel model} a variation of the magnetic field around a 
 resonance leads to a transition of the phase through $\pi/2$ and thereby
 drastically changes the value of $\ds a_{\rm sc}$. There are different types
 of normalization, \eg, the energy or momentum ones. For calculating
 observables like absolute transition rates the norm plays a role. However,
 general conclusions of the present work do not depend on the choice for the
 normalization. 

 The kinetic energy $E$ of two atoms when they are far apart is
 set to the arbitrarily chosen small value of 50\,Hz. Since this energy is
 very small, the collisions are limited to the 
 s-wave type only. The choice of a small but finite energy is justified
 because under ultracold conditions two particles collide with a low but
 non-zero energy. Furthermore, the non-zero energy helps to avoid non-physical
 numerical artifacts in the definition of the phase $\delta$.

 \subsection{Multi-channel results}
 \label{ssec:wavefunctions}

\begin{figure}[ht]
\centering
     \includegraphics[width=0.48\textwidth]{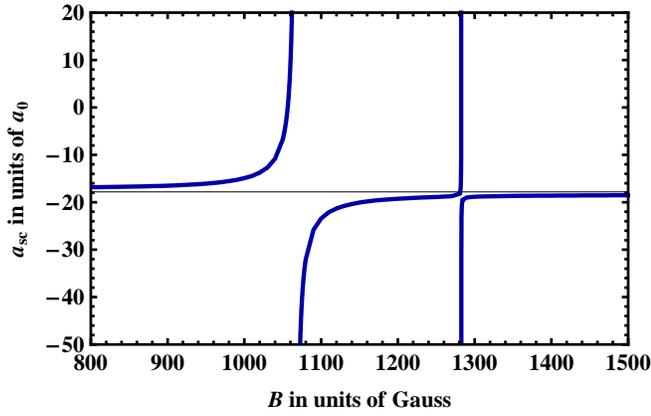}
 \caption{{\footnotesize
     (Color online)
     Scattering length $a_{\rm sc}$ as a function of the external magnetic
     field value $B$ for \LiRb scattering at $E=50\,$Hz. A broad and a narrow
     MFR are visible at $B_0=1066,917\,$G and $B_0=1282.576\,$G. The
     horizontal line marks the background scattering length 
     $a_{\rm bg} = -17.8\,a_0$ of the left resonance. 
}}
\label{fig:A_vs_B}
\end{figure}

 The system of \LiRb features for a collision energy $E=50\,$Hz two s-wave
 resonances in the range of $B<1500\,$G, a broad one at $B=1066,917\,$G and a
 narrow one at $B=1282.576\,$G (see Fig.~\ref{fig:A_vs_B}). 
\begin{figure}[!h]
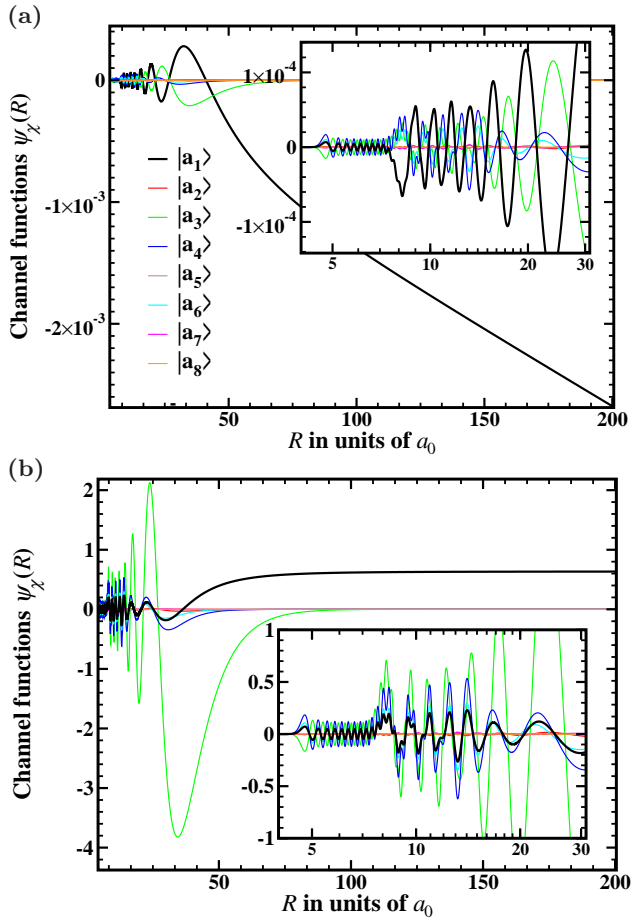

\centering
  {\bf (a)}\begin{minipage}[t]{0.46\textwidth}
     \vspace{0cm}\hspace{-1cm}
     \includegraphics[width=\textwidth]{eps/atombasisB0.eps}
     \end{minipage}
  {\bf (b)}\begin{minipage}[t]{0.46\textwidth}
     \vspace{0cm}\hspace{-1cm}
     \includegraphics[width=\textwidth]{eps/atombasisBr.eps}
     \end{minipage}
 \caption{{\footnotesize
     (Color online)
     The channel functions $\psi_{\chi}(R)$ for the 
     \LiRb collision in an off-resonant field $B=1000\,$G (a) 
     and a field $B=1066.9\,$G close to the resonance~(b). 
     The atomic labels (see Tab.~\ref{tab:atommolbasnote}) 
     are indicated in (a). 
     The insets focus on a region of small internuclear distance.
}}
\label{fig:atomic}
\end{figure}
\begin{figure}[!h]
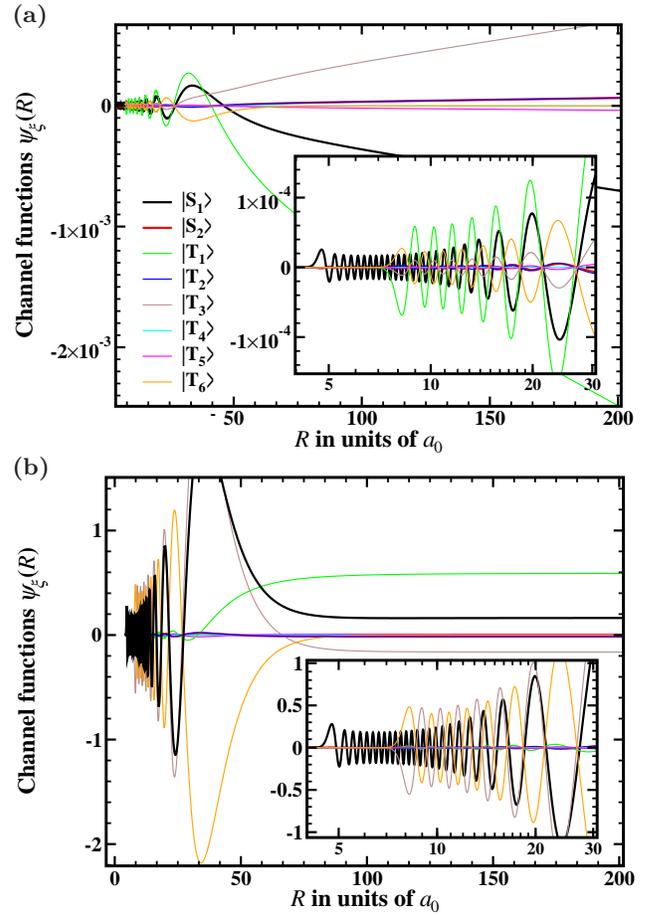

 \centering
  {\bf (a)}\begin{minipage}[t]{0.46\textwidth}
     \vspace{0cm}\hspace{-1cm}
     \includegraphics[width=\textwidth]{eps/molecbasisB0.eps}
     \end{minipage}
  {\bf (b)}\begin{minipage}[t]{0.46\textwidth}
     \vspace{0cm}\hspace{-1cm}
     \includegraphics[width=\textwidth]{eps/molecbasisBr.eps}
     \end{minipage}
 \caption{{\footnotesize
    (Color online)
     The channel functions $\psi_{\xi}(R)$. 
     The same as Fig.~\ref{fig:atomic} but in MB.
     The molecular labels are indicated in (a).
} }\label{fig:molecular}
 \end{figure}
 While the narrow resonance is also examined, this paper focuses on MC
 solutions around the broad resonance. This resonance has been also observed
 experimentally \cite{cold:deh08} and is well reproduced by the MC
 calculations. Moreover, processes like, \eg, PA are more efficient for a
 broad resonance because three-body losses can be minimized in this case. 

 Figures~\ref{fig:atomic} and~\ref{fig:molecular} present the channel functions
 of the MC calculations in AB and MB, respectively, for a collision of \LiRb at
 two different magnetic field strengths $B$. Figures.~\ref{fig:atomic}(a)
 and~\ref{fig:molecular}(a) show the case of a far off-resonant
 field of $B$=1000\,G which results in a small scattering length of only
 $a_{\rm sc}=-14.9\,a_0$. Figures~\ref{fig:atomic}(b)
 and~\ref{fig:molecular}(b) are taken close to the resonance at $B$=1066.9\,G
 with a scattering length of $a_{\rm sc}= -65\ 450\,a_0$. This large value
 is arbitrarily chosen for the present study. It is already a good
 representation of the resonant case $a_{\rm sc} =-\infty$. 

 The change of the long-range behavior between two scattering situations with
 small and large $a_{\rm sc}$ can be more clearly analyzed in the AB where all
 but one channel are closed, \ie, decay for large internuclear separations
 (see Fig.~\ref{fig:atomic}). As is evident from Figs.~\ref{fig:atomic}(a)
 and (b) the open-channel wave function $\psi_{a_1}$ changes the slope 
 resulting in a plateau when changing $a_{\rm sc}$ from small to
 large. Furthermore, the resonant open-channel function has a much larger
 amplitude within the considered range of interatomic distances than the
 off-resonant one. This large difference in amplitudes (about four orders of
 magnitude) sustains in the region of small internuclear distances (see insets
 of Figs.~\ref{fig:atomic}(a) and \ref{fig:atomic}(b)).

 At small internuclear distances above $R\approx7\,a_0$ the channel functions 
 in the AB show quite irregular behaviors (see insets of Fig.~\ref{fig:atomic}), 
 which is a result of the large coupling proportional to the exchange energy 
 $V_{\rm ex}(R)$. In the MB the coupling between the channels is induced by the 
 hyperfine interaction that is much smaller. Hence, the channel functions
 show a clear behavior of pure singlet and triplet wave functions for small
 internuclear distances (see insets of Fig.~\ref{fig:molecular}). For
 distances $R \leq 7\,a_0$ the triplet components vanish due to their higher
 exchange energy. Accordingly, also in the AB the channel functions are
 similar to pure singlet wave functions at $R \leq 7\,a_0$ (see insets of
 Fig.~\ref{fig:atomic}). All channel functions in the MB contribute
 correspondingly to the decomposition of the open channel $\psi_{a_1}$ into
 states of the MB. Therefore, at large internuclear distances they look
 similar to $\psi_{a_1}$. It is important to note that at small distances
 the closed channel functions have non-zero amplitudes even in the
 $B$-field-free case; they are slightly excited during the collision and
 possess a background contribution to the scattering process. Therefore, the
 two-body collision is a multi-channel process even in field-free space.

 Due to the resonant coupling at $B$=1066.9\,G, the admixture of the closed
 channels increases about four orders of magnitude. This is well described
 by the TC approximation~\cite{cold:fesh58,cold:koeh06} where the admixture of
 the closed channel and the long-range behavior of the open channel show a
 similar dependence on the scattering length 
 (see Sec.~\ref{ssec:Two-channel model}). In contrast to the TC approximation
 where one assumes that a bound state composed of a superposition of all
 closed channels is simply scaled at the resonance, the relative
 amplitudes change in reality between the resonant and off-resonant cases. On the other
 hand, the functional form of all closed channels indeed stays constant
 (compare, \eg, channel $\ket{a_4}$ in Figs.~\ref{fig:atomic}(a)
 and~\ref{fig:atomic}(b)). Altogether, this gives hope to be able to
 reproduce the change of the amplitude of both the open channel and the closed
 channels at small internuclear distances around an MFR with just one SC wave
 function.

 \subsection{Two-channel approximation}
 \label{ssec:Two-channel model}

 The TC approximation is very successfully used to describe resonance 
 phenomena in MC problems~\cite{cold:fesh58,cold:koeh06,cold:pell08}. 
 It is briefly introduced in order to understand to what extent SC approaches 
 can mimic MC systems. A more rigorous introduction may be found in
 \cite{cold:frie91,cold:marc04,cold:schn09a}.

 Within the TC approximation one projects the MC Hilbert space onto two subspaces,
 the one of the closed channels (with projection operator $\oper Q$) and the one of the 
 open channel (with projection operator $\oper P$). 
 The full wave function is thus written as $\ket{\Psi} = (\oper P + \oper Q)\ket{\Psi} = 
 \ket{\Psi_P} + \ket{\Psi_Q}$. 
 An MFR occurs, if the energy $E$ of the system is close to the eigenenergy
 $E_0(B)$ of a bound state $\ket{\Phi_b}$ of the closed-channel subspace.
 In the one-pole approximation one effectively assumes that the 
 closed-channel wave function is simply a multiple $A$ of the bound state $\ket{\Phi_b}$,
 \ie, $\ket{\Psi_Q}= A\, \ket{\Phi_b}$.
 This approximation yields the closed-channel admixture \cite{cold:schn09a}
\begin{equation}
  \label{eq:A2Ch}
  A = - \tilde{C} \sqrt{\frac{2}{\pi \Gamma}} \sin \delta_{\rm res}
\end{equation}
where $\tilde C$ is a normalization constant.
The long-range behavior of the open channel is given as
\begin{equation}
\label{eq:Asymp_OC}
 \left.\Psi_P(R)\right|_{R \rightarrow \infty} =
 \tilde C \sqrt{\frac{2 \mu}{\pi  k}} 
 \sin(k R + \delta_{\rm bg} + \delta_{\rm res})\,.
\end{equation}

 If the wave function is $\sin$ normalized, then $\tilde C =\sqrt{\frac{\pi k}{2 \mu}}$. 
 Another popular choice is the energy normalization with $\tilde C=1$. 
 However, the presence of an external trap can also induce a dependence of the
 normalization on the long-range behavior of the open channel parameterized by
 $a_{\rm sc}$, such that in general $\tilde C= \tilde C(a_{\rm sc})$.

 The total phase shift $\delta = \delta_{\rm bg} + \delta_{\rm res}$ results from the 
 background phase shift $\delta_{\rm bg}$ of the open channel without coupling to the closed channels
 and from a contribution $\delta_{\rm res}$ due to the resonant coupling to the bound state.  
 Via Eq.~(\ref{eq:delta_vs_scatlen}) the total phase shift is connected to the scattering length 
 $a_{\rm sc}$. The TC approximation yields for $k\rightarrow 0$ the well known
 relation~\cite{cold:moer95a}
 \begin{equation}
   \label{eq:a_vs_B}
   a_{\rm sc} = a_{\rm bg}\left(1 + \frac{\Delta B}{B - B_0} \right)
 \end{equation}
 between scattering length and magnetic field strength,
 where $a_{\rm bg} = -\tan{\delta_{\rm bg}}/k$ is the background scattering length, $\Delta B$ is the 
 width of the resonance, and $B_0$ its position.

 We note that independently of the normalization function 
 $\tilde C(a_{\rm sc})$ both the admixture of the closed channel $A$ and the long 
 range open-channel solution~(\ref{eq:Asymp_OC}) show for small energy, not too large
 internuclear distances (\ie, $k R \ll \delta$), 
 and small background phase shifts (\ie, $\delta\approx\delta_{\rm res}$)
 a similar dependence on the scattering length $a_{\rm sc}$. 

 Usually, for small energy $E$ the background phase shift 
 $\delta_{\rm bg} = -\arctan(k a_{\rm bg})$ is necessarily also small. 
 Since a scaling of the open-channel wave function in the long range is more or less 
 directly continued to shorter distances, the proportionality between $A$ and $\Psi_P(R)$
 holds approximately also for smaller $R$. 
 Therefore, looking at molecular processes taking place at small internuclear 
 distances, the enhancement of the closed-channel contribution is already
 reproduced by the open channel. This paves the way to an SC description which
 will now be discussed.

 \section{Single-channel approaches}
 \label{sec:SC_appr}

 \subsection{Variations of the single-channel Hamiltonian}
 \label{sec:SC_Ham}

 In order to reflect the molecular behavior at small distances, we will seek to 
 base the SC approximations on pure singlet or triplet interaction 
 potentials. This ensures that the
 nodal structure of the resulting SC wave function is similar
 to the relevant singlet or triplet components of the MC system.
 The final aim is to mimic in parallel the long-range behavior of the open
 channel and the variation of the amplitude of singlet or triplet components
 in the vicinity of an MFR.

 In an SC approach the interaction strength can be artificially varied by a controlled 
 manipulation of the Hamiltonian 
 \begin{equation}
   \ds H(R) =
      -\frac{1}{2\mu}\frac{\partial^2}{\partial R^2}+V(R)\,.
       \label{eq:HSC}
 \end{equation}
 Subject to modification are the inter-atomic potential $V(R)$ and the reduced mass
 $\mu$ of the system. The modifications can lead to a shift of the energy of the least
 bound state relative to the potential threshold.
 When lifted above the threshold, the bound state turns into a virtual 
 state~\cite{cold:newt02,cold:marc04}. A large scattering length of the
 solution of the SC Schr\"odinger equation with Hamiltonian~(\ref{eq:HSC}) can
 be elegantly explained by a resonance of the scattering state with either a
 real bound state or a virtual state close to the
 threshold~\cite{cold:newt02,cold:marc04}. Within an SC approach the energy
 of a bound or virtual state is changed in order to induce a variation of the
 scattering length. In this respect SC approaches show striking similarities
 to MFRs where the energy of a bound state in the closed-channel subspace is
 moved by changing its Zeeman energy by an external magnetic field.

 As argued before, the SC wave functions should be either of singlet or
 triplet character for small internuclear distances. We reduce our
 considerations for Li-Rb to the singlet case and chose as initial potential 
 the one for the $X^1\Sigma^+$ electronic ground state, \ie,
 $V(R)=V_{X^1\Sigma^+}(R)$. This potential is varied by a controlled
 manipulation of the strong-repulsive inner wall~\cite{cold:gris09}, 
 the long-range van der Waals attraction $V_{\rm vdW}(R)$~\cite{cold:ribe04}, 
 and a novel Gaussian perturbation around the transition point $R_{\rm sh}$
 between the molecular and the atomic description of the system (introduced in
 Sec.~\ref{subsubsec:molecbas}). These procedures will be called $s$ variation,
 $C_6$ variation, and $G$ variation, respectively.

\begin{figure}[!ht]
 \centering
 \includegraphics[width=0.46\textwidth]{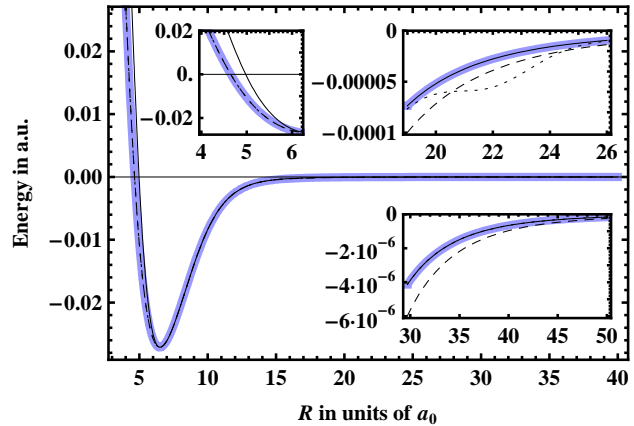}
 \caption{{\footnotesize
     (Color online)
      Original $X^1\Sigma^+$ potential $V(R)$ (blue solid) with applied 
      s variation (black solid), $C_6$ variation (dashes) and $G$ variation
      (dots). The variation parameters are $s=0.03\,a_0$, $\dC6 = C_6/2$
      and $\dG = V(R_{\rm sh})$ (see Tab.~\ref{tab:VarPar} for the adopted
      parameters). The insets show some relevant ranges of $R$ on an enlarged
      scale. 
    }}\label{fig:PotVariations}
 \end{figure}

 The potential variations are induced by replacing $V(R)$ by
 \begin{eqnarray}
   \ds   &\,&V^s(R) = 
        \begin{cases}
          V(R-s\cdot\frac{R-R_e}{R_c-R_e})  &  R\leq R_e \\
          V(R)  &  R>R_e
        \end{cases}
        \,,\\
        &\,& V^{\dG}(R) = V(R) + 
        \dG\, {\rm Exp}{\left(\frac{R-R_G}{\sigma}\right)^2}\,,
\end{eqnarray}
 or
 \begin{eqnarray}
    V^{\dC6}(R) = V(R) + \frac{\dC6}{R^6}\cdot f(R)
 \end{eqnarray}
 where $R_e=6.5\,a_0$ is the equilibrium distance and $R_c=4.6\,a_0$ is the 
 crossing point of the $V_{X^1\Sigma^+}(R)$ with the threshold. The width in
 the $G$ variation is chosen as $\sigma=2\,a_0$ and its position as
 $V_G=R_{\rm sh}+\sigma$. The smooth variation of the long-range region of the
 potential in the $C_6$ variation is achieved by the gradual stepping function
 \begin{equation}
   \label{eq:fud}
   \ds 
   f(R) = \left(1+e^{\frac{\gamma(R_0-R)}{\Delta}}\right)^{-1}
 \end{equation}
 where $\gamma = \ln(999) \approx 6.9$ ensures that $f(R)$ rises from 
 0.001 to 0.999 in the region $R_0-\Delta\leq R \leq R_0+\Delta$. 
 For the present study the parameters of the tuning function are chosen as
 $\Delta=6$ and $R_0=16 a_0$.  The three potential variations are depicted 
 in Fig.~\ref{fig:PotVariations}.

 An alternative way to tune $a_{\rm sc}$ is offered by the $\mu$ variation
 within which one changes the reduced mass of the system by 
 $\mu \rightarrow \mu - \dm$~\cite{cold:gris07}.  
 This alters the kinetic energy operator and can modify the energy of 
 the least bound state like potential variations. In contrast to the presented
 potential variations which act on either the short-range, mid-range or
 long-range part of the potential, the mass variation influences the
 Schr\"odinger equation at any distance. It is very similar to a scaling of
 the potential by $V(R)\rightarrow \gamma V(R)$ \cite{cold:esry99a}. The only
 difference is an additional change of the energy-momentum relation 
 $E(k)= k^2/(2\mu)$ which can, \eg, slightly influence the normalization of
 the wave function. 

 One can think of several other approaches to vary the SC Hamiltonian.
 For example, one can vary the strength of the exchange energy $J_0$, its decay
 parameter $\alpha$ or the van der Waals parameters $C_8,C_{10}$ \cite{cold:ribe04,cold:ribe04a}. 
 The current approaches are chosen to comprise variations that act on the short rage 
 of interatomic distance ($s$ variation), on an intermediate range ($G$ variation), 
 on the long range ($C_6$ variation), and on the full range ($\mu$ variation).

 No matter which SC approach is finally chosen, a mapping between the MC system and an 
 appropriate SC Hamiltonian is straightforward.
 Knowing the parameters $\Delta B$ and $B_0$ in Eq.~(\ref{eq:a_vs_B}) for an MFR either from 
 experimental data or a coupled MC calculation one can connect each value of the magnetic field
 $B$ to a scattering length $a_{\rm sc}$ and a corresponding value of the SC variation parameter 
 that induces the same value of the scattering length.
 Clearly, this additional information is required, \ie, the SC model has no predictive power by 
 itself.

 Typical values of the four variation parameters as they will be used in the 
 following are given in Tab.~\ref{tab:VarPar}.
 The wave functions resulting from the different variation methods are 
 denoted $\phi^{\upsilon}(R)$ where $\upsilon\in\{s, G, C_6, \mu\}$ stands for 
 the applied $\upsilon$ variation. 
\begin{table}
  \caption{Values of the parameters for $s,C_6,G,\mu$ variations
    at small and large scattering lengths resulting from a magnetic 
    field far away, close and right at an MFR. An infinitesimally small 
    change of the field right at the resonance ($B=1066.92\,a_0$) 
    switches the interaction regime from the infinitely attractive 
    ($a_{\rm sc}=-\infty$) to infinitely repulsive 
    ($a_{\rm sc}=+\infty$).} 
  \begin{ruledtabular}
     \begin{tabular}{cccccc}
         $B/$Gauss  & $a_{\rm sc}/a_0$ & $s/a_0$  & $\dC6/C_6$  & $\dG/|V(R_{\rm sh})|$& $\dm/\mu$  \\
         \hline
         1000.00    &  -14.93          & -0.00947 &  -0.0281    &  -0.1310             & -0.00235\\

\vspace{0.1cm}
         1066.90    &  -65450          & -0.04142 &  -0.1282    &  -0.6296             & -0.01046\\
	\multirow{2}{*}{1066.92}  & $-\infty$
                                       & -0.04145 &  -0.1283    &  -0.6302             & -0.01047\\
                                  & $+\infty$   &  0.13065 &   0.3357    &  -4.3856             &  0.02984\\
     \end{tabular}
  \end{ruledtabular}
  \label{tab:VarPar}
\end{table}

 \subsection{Multi-channel vs single-channel}
 \label{subsec:SCMCwfcomp}

\begin{figure}[!ht]
 \centering
 \includegraphics[width=0.46\textwidth]{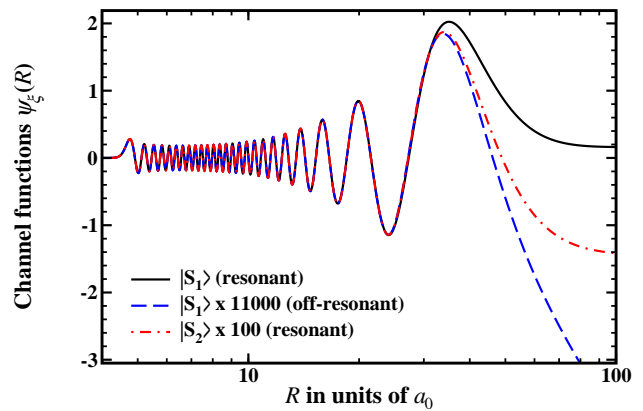}
 \caption{{\footnotesize
     (Color online)
     The channel functions of the singlet state 
     $\ket{S_1}$ close to the resonance ($a_{\rm sc}=-65\ 450\,a_0$) and
     away from the resonance ($a_{\rm sc}=-14.9\,a_0$) are depicted together with
     the channel function of the close to resonant $\ket{S_2}$ state. 
     All three functions differ for $R<30\,a_0$ only by a constant prefactor.
    }}\label{fig:singetscompare}
 \end{figure}

 In the following, the wave functions of the SC and MC approaches are
 compared. As discussed in Sec.~\ref{subsubsec:molecbas}, the appropriate
 choice of the MC basis depends on the interatomic distance. While for large
 interatomic distances ($R>R_{\rm sh}$) the description in the AB is adequate,
 their basis states are strongly coupled for shorter distances. Here, the MB
 describes the physical properties far better. Due to the weak hyperfine
 coupling the states in the MB keep to a good degree of accuracy the structure
 of an uncoupled singlet or triplet state, respectively. Close to an MFR
 solely amplitudes for some of the states are heavily increased. Figure~\ref{fig:singetscompare}  
 shows for example a comparison of the singlet state $\ket{S_1}$ close to an
 MFR with the same state far away from the resonance and with the other
 singlet state $\ket{S_2}$ again close to the resonance. Clearly, for
 $R<R_{\rm sh}$ they differ only by a constant prefactor. This is important,
 as it allows to describe the short-range behavior of the MC wave function by
 either a pure singlet or triplet channel function depending on the physical
 process which is to be described. For example, only singlet components
 contribute to the DPA process for the transition into the absolute
 vibrational ground state (as it will be discussed in Sec.~\ref{sec:PAtoGS}), hence,
 the triplet components may be omitted. 

 In the following, the aim of the SC approach is to mimic the behavior 
 of the MC singlet components for $R<R_{\rm sh}$ by a controlled variation of
 the SC Hamilton operator~(\ref{eq:HSC}) with singlet potential
 $V_{X^1\Sigma^+}(R)$. With the help of the $s,C_6,G$ and $\mu$ variations presented
 in Sec.~\ref{sec:SC_Ham} the SC wave function is adjusted to match the
 asymptotic behavior (\ie, the scattering length $a_{\rm sc}$) of the open
 channel for a given external magnetic field $B$. The cases of an off-resonant
 magnetic field ($B=1000.0\,$G) and one close to a resonance $(B=1066.9\,$G)
 are considered. The corresponding scattering lengths are 
 $a_{\rm sc}=-14.9\,a_0$ and $a_{\rm sc}=-65\ 450\,a_0$ (see
 Sec.~\ref{ssec:wavefunctions} and Figs.~\ref{fig:molecular},~\ref{fig:atomic}). 
 \begin{figure}[!ht]
 \centering
  {\bf (a)}\begin{minipage}[t]{0.48\textwidth}
     \vspace{0cm}\hspace{-1cm}
     \includegraphics[width=\textwidth]{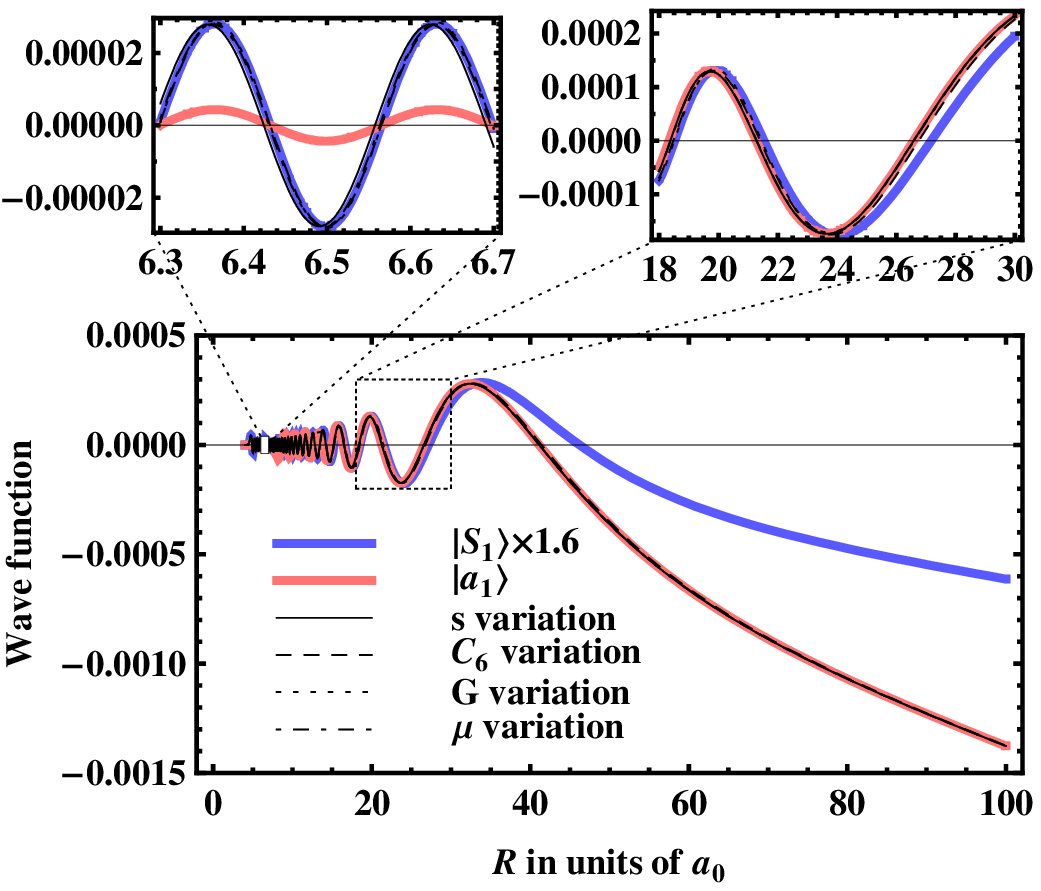}
     \end{minipage}
  {\bf (b)}\begin{minipage}[t]{0.48\textwidth}
     \vspace{0.2cm}\hspace{-1cm}
     \includegraphics[width=\textwidth]{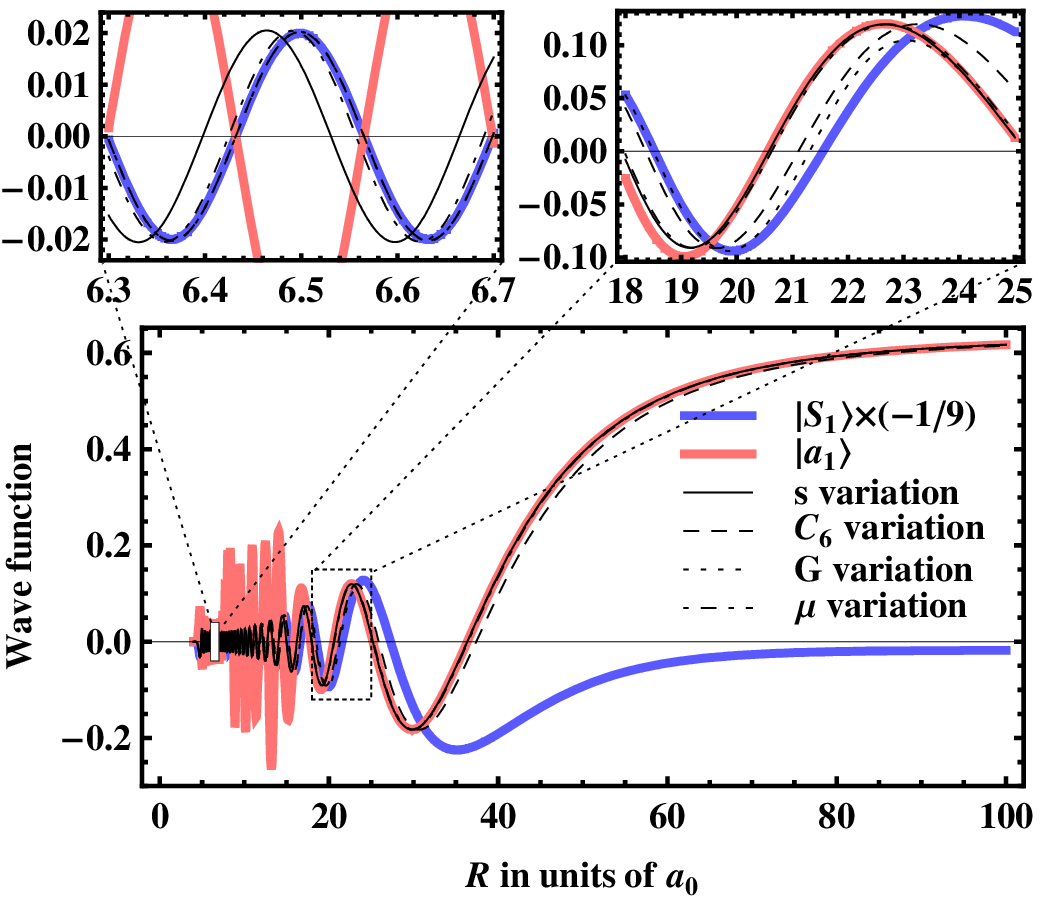}
     \end{minipage}
 \caption{{\footnotesize
     (Color online)
     Comparison of the SC wave functions $\phi^\upsilon(R)$
     with the MC functions of the singlet state $\ket{S_1}$ (scaled) and 
     the open channel $\ket{a_1}$.
     The SC potentials are varied to match the asymptotic behavior 
     of the MC channel functions of the open channel.
     {\bf (a)} Off-resonant case with $a_{\rm sc}=-14.9\,a_0$ ($B=1000\,$G).
     {\bf (b)} Resonant case with $a_{\rm sc}=-65\ 450\,a_0$ ($B=1066.9\,$G).
     The according values of the $s,\dC6,\dG$, and $\dm$ parameter 
     are given in Tab.~\ref{tab:VarPar}.
     The smaller plots focus respectively on a region of small internuclear 
     distance (left) and a region $R\approx R_{\rm sh}$ (right).
 } }\label{fig:Comp_SC_MC}
 \end{figure}

 Figure~\ref{fig:Comp_SC_MC} shows a comparison of the SC wave functions
 $\phi^\upsilon(R)$ with the channel functions of the dominant singlet channel
 $\ket{S_1}$ and the open channel $\ket{a_1}$ for the full range of short and
 long interatomic distances. Figure~\ref{fig:Comp_SC_MC} allows to examine how
 the different variational methods are able to reflect both the behavior of
 the singlet components for distances $R<R_{\rm sh}$ and the one of the open
 channel for $R>R_{\rm sh}$.

 Generally, any SC approach has to induce a shift of the phase $\delta$ in order 
 to tune the scattering length $a_{\rm sc}=-\tan(\delta)/k$. The difference 
 $\delta-\delta_{\rm ini}$ from the phase of the unperturbed system $\delta_{\rm ini}$ 
 is accumulated where the variation of the SC Hamiltonian takes place.
 Since the scattering length of the original singlet potential $V_{X^1\Sigma^+}(R)$ 
 is with $a_{\rm sc}^{\rm ini}=2.3\,a_0$ relatively close to 
 $a_{\rm sc}=-14.9\,a_0$, hardly any phase shift has to be acquired
 ($\delta-\delta_{\rm ini} = 8.6\cdot 10^{-5}\pi$) to match 
 the open channel for the off-resonant magnetic field $B=1000\,$G. 
 Accordingly, the nodal structure of the MC singlet component
 is very well matched in Fig.~\ref{fig:Comp_SC_MC}(a).
 The situation changes close to the resonance where the large scattering length 
 $a_{\rm sc}=-65\ 450\,a_0$ requires a phase shift of $\delta-\delta_{\rm ini} = 0.22\pi$
 (Fig.~\ref{fig:Comp_SC_MC}(b)).
 This is about half way to the resonant phase shift $\pi/2$.

 For the $s$ variation the total phase shift to match the open channel is acquired
 for distances $R<6.5\,a_0$. Accordingly, the nodal structure between the $\ket{S_1}$
 channel function and the SC $\phi^s(R)$ wave function is shifted for $R>6.5\,a_0$ 
 (see upper left plot in Fig.~\ref{fig:Comp_SC_MC}(b)).
 Contrarily, both the $C_6$ variation and the $G$ variation induce a phase shift for 
 distances $R$ larger than $R_G-\sigma = 20\,a_0$ and $R_0=16\,a_0$.
 Thus, for smaller internuclear distances, the SC wave functions $\phi^G(R)$
 and $\phi^{C_6}(R)$ coincide with the $\ket{S_1}$ channel function. Finally,
 since the $\mu$ variation acts on any internuclear distance, the phase
 difference is gradually accumulated for $\phi^{\mu}(R)$.

 Depending on the range of variation for the SC approaches, also the matching to the
 open channel of AB differs. The $s$ variation matches the open channel already closely 
 before $R_{\rm sh}$. Surprisingly, also the $\mu$ variation shows a reasonable match 
 already before $R_{\rm sh}$, although it acts also for larger distances by changing at
 least the dispersion relation $E(k)$. This effect may, however, not be visible, since 
 $k R \ll 1$ in the plotted region.
 The $C_6$ variation changes the long-range behavior of the interaction potential. 
 Correspondingly, the wave function shows a clear difference to the open channel even 
 up to $R=100\,a_0$. 
 The Gaussian perturbation of the G variation acts only around $R_{\rm sh}$.
 This results in the favorable situation that both the $\ket{S_1}$ channel
 for $R<20\,a_0$ and the open channel for $R>24\,a_0$ are matched by the SC
 wave function.

 Since the nodal structure among different singlet and different triplet channels
 coincides for $R<R_{\rm sh}$ the presented results are generalizable to any 
 singlet or triplet state. Thus, SC approaches are generally 
 able to reproduce the asymptotic behavior of the open channel of the MC wave function 
 in the presence of an MFR while also reflecting certain aspects of singlet or triplet 
 components for small internuclear distances. Depending on the region of the variation of the
 SC Hamiltonian, the nodal structure of any channel function in the MB can be reproduced
 for $R<R_{\rm sh}$. The most flexible SC approach 
 is the $G$ variation which is able to smoothly switch between the accurate description
 of a MB channel and the open channel.  Furthermore, it offers the advantage, that one
 can define the transition point (here $R=R_{\rm sh}$) at will, such that
 also for slightly larger distances MB channel functions can be emulated.

 An aspect of the MB channels which cannot be reflected by the present approaches 
 is their absolute amplitude. Since the amplitudes at small internuclear distances 
 of the different channels change drastically in the presence of an MFR, 
 they have a large impact on molecular processes such as the association
 of molecules utilizing MFRs. In the next section the exemplary case of a direct
 dumping of the scattering state to the vibrational ground state of the $X^1\Sigma^+$
 is considered. The transition rate depends strongly on the behavior of the amplitude
 of the dominant singlet state $\ket{S_1}$  which was considered in this section. 
 It will be shown that although the absolute amplitude of this state is not reproduced
 by any SC approach, the relative enhancement of the transition rate at magnetic
 fields close to a resonance can be well reflected.

 \section{Photoassociation of {\LiRb} to the absolute vibrational ground state}
 \label{sec:PAtoGS}

 Ultracold polar molecules are of great interest for many applications in quantum 
 information processing \cite{cold:mich06, cold:rabl06}, the exploration of lattices of 
 dipolar molecules \cite{cold:pupi08}, precision measurement of fundamental constants
 \cite{cold:zele08}, and ultracold chemical reactions \cite{cold:chin05,cold:tsch06}.
 Since standard cooling technics developed for atoms are not suitable for molecules 
 due to their complex level structure, ultracold molecules may alternatively be
 associated directly from ultracold atoms.
 As was already mentioned in the introduction the starting point to create
 ultracold molecules in their vibrational ground state are often Feshbach molecules
 formed by a sweep of the magnetic field around an MFR in a high-lying
 vibrational level~\cite{cold:koeh06}. These loosely bound molecules are
 usually transferred by complex PA schemes via intermediate excited states to
 the desired vibrational ground state~\cite{cold:ni08,cold:sage08}. Especially
 STIRAP~\cite{cold:berg98,cold:wink08,cold:ospe08,cold:danz08} showed to
 be successful in efficiently creating ultracold ground state molecules. 
 However, Feshbach molecules possess a relatively short life time such that a
 Feshbach optimized transition directly at the resonance can be
 favorable~\cite{cold:kuzn09}.   

 For all schemes that make advantage of the resonant coupling to a molecular bound state 
 at an MFR~{\cite{cold:abee98,cold:cour98,cold:rega03,cold:gris07,cold:junk08,cold:pell08,cold:deig09}}, 
 the increase of the amplitude for the relevant channels as 
 the scattering length grows is of great importance to enhance the molecule creation. 
 Although in the last section it was shown that the absolute
 amplitude of the MB channels is not reproduced by the SC approaches, the TC approximation gives 
 hope that the relative enhancement can still be recovered. In 
 Sec.~\ref{ssec:Two-channel model} it was discussed that both the admixture of the closed-channel 
 bound state and the open-channel function scale similarly with the 
 scattering length. One can therefore expect to be able to combine this collective relative enhancement 
 into one channel.

\begin{figure}[!ht]
 \centering
     \includegraphics[width=0.46\textwidth]{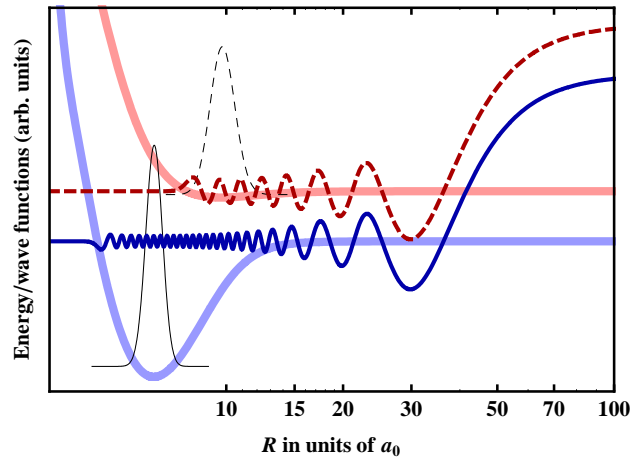}
 \caption{{\footnotesize
     (Color online)
     Sketch of the resonant SC wave functions obtained via $G$ variation 
     and respective BO potentials relevant for the DPA transition to either
     the singlet ground state (thin, solid) or the triplet ground state (thin,
     dashed) of the respective $X^1\Sigma^+$ (thick, blue) and $a^3\Sigma^+$
     (thick, red) potentials.
     For better visibility, the potentials and wave functions
     are shifted along the y-axis. In reality, singlet and triplet potentials 
     have the same threshold energy.
 } }\label{fig:transition}
 \end{figure}

 In the following, the Feshbach optimized DPA (FOPA)~\cite{cold:kuzn09} to the absolute vibrational ground  
 state of \LiRb in the electronic  $X^1\Sigma^+$ state is considered to examine the applicability of 
 SC approaches to study processes of molecule creation. We consider this case since it has an interest on 
 its own for the creation of bound ultracold molecules. Furthermore, the transition rate to
 the absolute vibrational ground state depends on
 the scattering wave function at very small internuclear distances (see Fig.~\ref{fig:transition}). 
 We also examined the transition to the vibrational ground state of the electronic triplet state 
 $a^3\Sigma^+$ which is situated at slightly larger interatomic distances. Since we found no essential
 differences to the singlet case, we focus on presenting only its results in this work.

\subsection{Calculation of transition rates}

 Given the solution of the MC problem $\Psi(R) = \sum_{\xi}\frac{\psi_\xi(R)}{R}\ket{\xi}$ 
 in the MB for a given magnetic field $B$ the free-bound FOPA transition rate 
 $\Gamma_{\downarrow}(B)$ to the final molecular state $\Psi_{\rm f}(R) =
 \frac{\Psi_{\nu}(R)}{R}Y_J^M(\Theta,\Phi) \ket{\xi_{\rm f}}$ with 
 vibrational quantum number $\nu$ and rotational quantum number $J$
 within the dipole approximation is proportional to the squared dipole transition moment
 \cite{cold:sand71}
 \begin{equation}
   \ds
   \label{eq:Iv}
   I_{\rm MC}(B) = 
            \left|
            \int\limits_0^{\infty}
            \Psi_\nu(R)
            D(R)
            \psi_{\xi_{\rm f}}(R) dR
            \right|^2\,.
 \end{equation}
 Here, $D(R)$ is the electronic dipole moment.
 Within the dipole approximation only transitions from the s-wave scattering function
 to a final state with $J=1$ are allowed.
 Due to the orthogonality of the MB, only one molecular channel has to be taken into account
 in Eq.~(\ref{eq:Iv}).

 The TC approximation predicts a rate \cite{cold:schn09a}
 \begin{equation}
  \ds
   \label{eq:IvSimp2Ch}
   I_{\rm TC}(B) =  |\tilde C \cdot \mathcal C|^2 \cdot 
    \left|\sin \left(\delta_{\rm res}(B) - \delta_0 \right)\right|^2
 \end{equation}
 where the constants $\mathcal C$ and $\delta_0$, explicitely given in \cite{cold:schn09a},
 do not depend on the magnetic field within the TC approximation. The phase shift $\delta_0$ is
 usually small \cite{cold:schn09a} and thus the minimum lies close to a vanishing 
 resonant phase shift $\delta_{\rm res}=0$, \ie, close to the background scattering length 
 $a_{\rm bg}$.
 We determine $a_{\rm bg}$ by a fit of $a_{\rm sc}(B)$ to Eq.~(\ref{eq:a_vs_B}) which 
 yields $a_{\rm bg}=-17.8 a_0$. From $\delta = \delta_{\rm res} + \delta_{\rm bg}$
 and Eq.~(\ref{eq:a_vs_B}) one can then directly determine $\delta_{\rm res}(B)$.
 The behavior of Eq.~(\ref{eq:IvSimp2Ch}) accurately reflects
 the one of a MC system for well separated resonances \cite{cold:schn09a}.
 We use it here to determine the maximal MC transition rate.

 The transition rate $\Gamma_{\downarrow}^{\upsilon}$ to the final state within an SC approach 
 is simply proportional to
 \begin{equation}
   \label{eq:IvSC}
 I_{\rm SC}^{\upsilon}(a_{\rm sc}) = 
 \left|\int_0^\infty \Psi_{\nu}(R) D(R) \phi^\upsilon(R) \right|^2 
 \end{equation}
 where $\upsilon$, as before, denotes the variational method which for the
 present analysis induces the scattering length $a_{\rm sc}$ equal to the one
 of the MC system for a given $B$-field value. 

 $D(R)$ is again the electronic dipole transition moment. For the purpose of
 the present study we reduce our considerations to the linear approximation
 $D(R) = D_0 + D_1\cdot R$. The SC scattering wave function is orthogonal to
 the different vibrational bound states. In the MC case only the weak
 hyperfine coupling in the MB causes a very slight non-orthogonality. The
 influence of $D_0$ can be therefore safely ignored. Calculations with
 higher-order expansions showed that the exact functional behavior of $D(R)$
 (obtainable from~\cite{cold:ayma05}) does hardly influence the relative
 enhancement of the transition rate. Thus, the use of $D(R)=D_1\cdot R$  
 does not restrict generality. It is important to note that
 Eqs.~(\ref{eq:Iv})-(\ref{eq:IvSC}) are only valid within the dipole
 approximation. It is supposed to be applicable, if the wavelength of the
 associating photon is much larger than the spatial extension of the atomic or
 molecular system. The shortest PA laser wavelength corresponds to the
 transition to the lowest vibrational state. Although the spatial extention of
 the initial state is infinite, the integrals for dipole transition moments is
 finite, as it contains a finite wave function of the bound vibrational state
 as a factor. Therefore the dipole approximation is valid.

\subsection{Comparison of transition rates}

 A change of $a_{\rm sc}$ leads to an increase or decrease of
 $\Gamma_{\downarrow}^v$. In order to quantify the magnitude of this change, an
 enhancement or suppression factor may be introduced~\cite{cold:gris07}
 \begin{equation}
   g^v(a_{\rm sc}) =
   \frac{\Gamma_{\downarrow}^v(a_{\rm sc})}
        {\Gamma_{\downarrow}^v(a_{\rm sc}^{\rm ref})}
                   =
   \frac{I^v(a_{\rm sc})}
   {I^v(a_{\rm sc}^{\rm ref})} \quad .
   \label{eq:gv}
 \end{equation}
 It describes the relative enhancement [$g^v>1$] or suppression [$g^v<1$] of
 the DPA rate at a given $a_{\rm sc}$ vs. a~reference scattering length 
 $a_{\rm sc}^{\rm ref}$, for a specific final state $v$. Although it may
 appear to be most natural to choose $a_{\rm sc}^{\rm ref}=0$, a large
 non-zero value offers some advantages. In this case, 
 $I^v(a_{\rm sc}^{\rm ref})$ is not too small and large numerical
 errors are avoided.

\begin{figure}[!h]
 \centering
  {\bf (a)}\begin{minipage}[t]{0.48\textwidth}
     \vspace{0cm}\hspace{-1cm}
     \includegraphics[width=\textwidth]{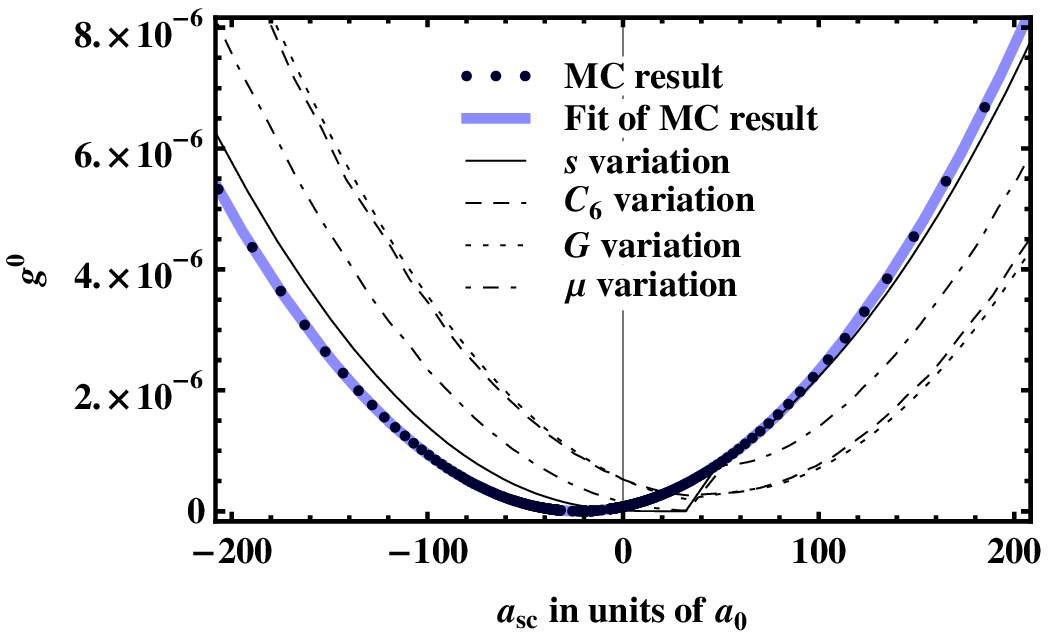}
     \end{minipage}
  {\bf (b)}\begin{minipage}[t]{0.48\textwidth}
     \vspace{0.2cm}\hspace{-1cm}
     \includegraphics[width=\textwidth]{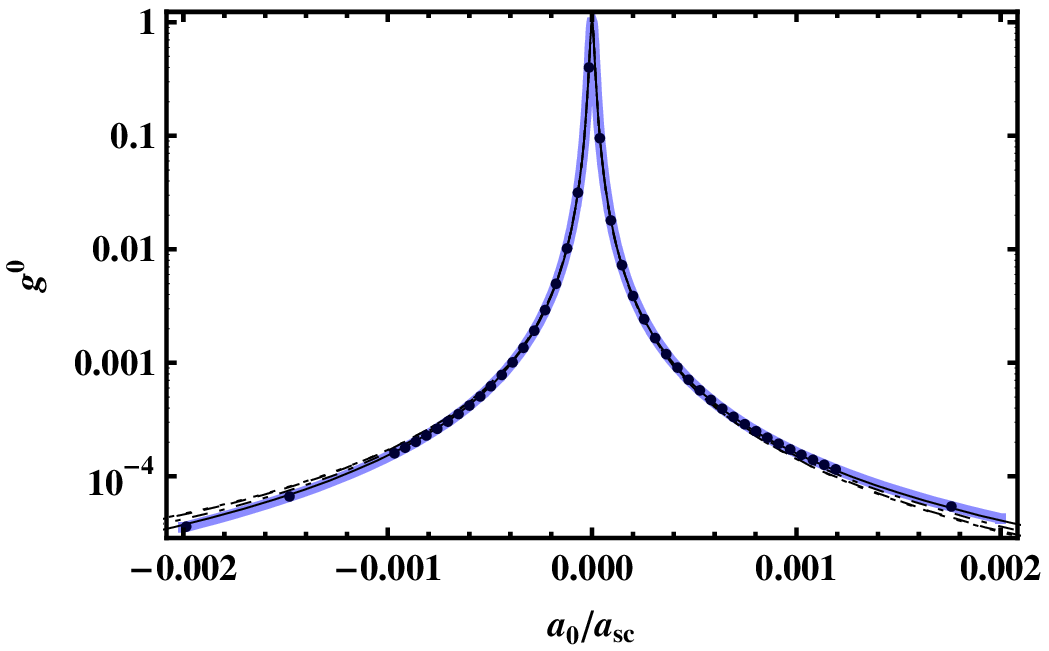}
     \end{minipage}
 \caption{{\footnotesize
     (Color online)
     Comparison of MC and SC results for the transition rate to the 
     absolute vibrational ground state relative to the respective 
     maximal transition rate as a function of the scattering 
     length (a) and the inverse scattering length (b).
     The MC results are fitted according to the 
     TC approximation (Eq.~(\ref{eq:IvSimp2Ch})).
 } }\label{fig:Comp_Rate_SC_MC}
 \end{figure}

 Figure~\ref{fig:Comp_Rate_SC_MC} shows a comparison of the SC transition rate for the 
 different  variational approaches with the correct MC result. 
 In the calculation of the MC transition rates, we assume a measurement in
 which the nuclear spins are not resolved. This corresponds in practice to the
 case in which the transition rates from the $\ket{S_1}$ and $\ket{S_2}$
 channel are summed. All rates are normalized to their respective maximum
 value ($a_{\rm sc}^{\rm ref}=\infty$). Note, however, that the different
 absolute dipole transition moments disagree by some orders of magnitude.

 A fit of the MC result by the TC 
 estimate with only two free parameters $\mathcal C$ and $\delta_0$ reveals that 
 the simple dependence of the transition rate
 given by Eq.~(\ref{eq:IvSimp2Ch}) describes the transition process of the MC system
 correctly. 

 All SC approaches agree with the MC result for large scattering lengths 
 in the proximity of the resonance (see Fig.~\ref{fig:Comp_Rate_SC_MC}(b)).
 For small scattering lengths where the transition rate is already suppressed by
 more than four orders of magnitude deviations from the MC result appear. The differences
 mainly originate from a shift of the minimal transition rates of the SC approaches compared to 
 the MC result.
 In the MC case the minimum lies at $a_{\rm sc}=-21.1 a_0$ close to the background scattering 
 length $a_{\rm bg} = -17.8 a_0$ in accordance with the TC approximation. The
 minima of the SC approaches tend to be situated on the positive side around 
 $a_{\rm sc}\approx 50 a_0$. This is, however, not a general trend, since we observed
 for other transitions also minimal SC transition rates at negative scattering
 lengths. The location of the minimum depends on a system under investigation
 and on the applied SC variation. 

 Figure~\ref{fig:Comp_Rate_SC_MC}(a) features two kinks of the transition rate at 
 $a_{\rm sc}=40\,a_0$ for the $s$ and $\mu$ variations. This can be explained by the shift 
 of the nodes of the SC wave functions which takes place at the equilibrium distance of the 
 bound molecule and therefore influences the PA rate. Since 
 the variation parameters are tuned around their resonance value,
 with increasing distance from the resonance both left and right of it, eventually the 
 same scattering length is induced (see Fig.~\ref{fig:aswf}). However, the nodal structure  
 of $\phi^s(R)$ and $\phi^\mu(R)$ for short ranges can differ, leading to different transition 
 rates. This does not occur for the $C_6$ and $G$ variations that act far beyond the 
 equilibrium distance. Note, however, the scale at which the kink is
 visible. Its effect on the rate is minute.

 Analogous examinations were also done for the other MFR of \LiRb at 
 $B=1282.58\,$G. Although this resonance is two orders of magnitude narrower
 than the one considered before and the amplitudes of the channels are different,
 no significant differences for the relative rates were observed. The
 generality of our considerations is also supported by calculations of the
 dumping rate to the vibrational ground state of the triplet configuration
 $a^3\Sigma^+$. In all cases the SC approaches showed a comparable ability to
 reflect results of the MC system. 

 It is also interesting to note that results of the $g^0$
 analysis show that neither the details of the interatomic nor magnetic-field
 interactions are relevant for the calculation of the relative rate. 
 A simple SC model turns out to be adequate to calculate the relative
 enhancement of the PA process. Furthermore, in view of the important question
 of how to optimize the efficiency of DPA, Fig.~\ref{fig:Comp_Rate_SC_MC}
 reveals once more that the use of a large absolute value of the scattering
 length is favorable.

\subsection{Number of bound states}
 
\begin{figure}[!ht]
 \centering
  {\bf (a)}\begin{minipage}[t]{0.45\textwidth}
     \vspace{0cm}\hspace{-1cm}
     \includegraphics[width=\textwidth]{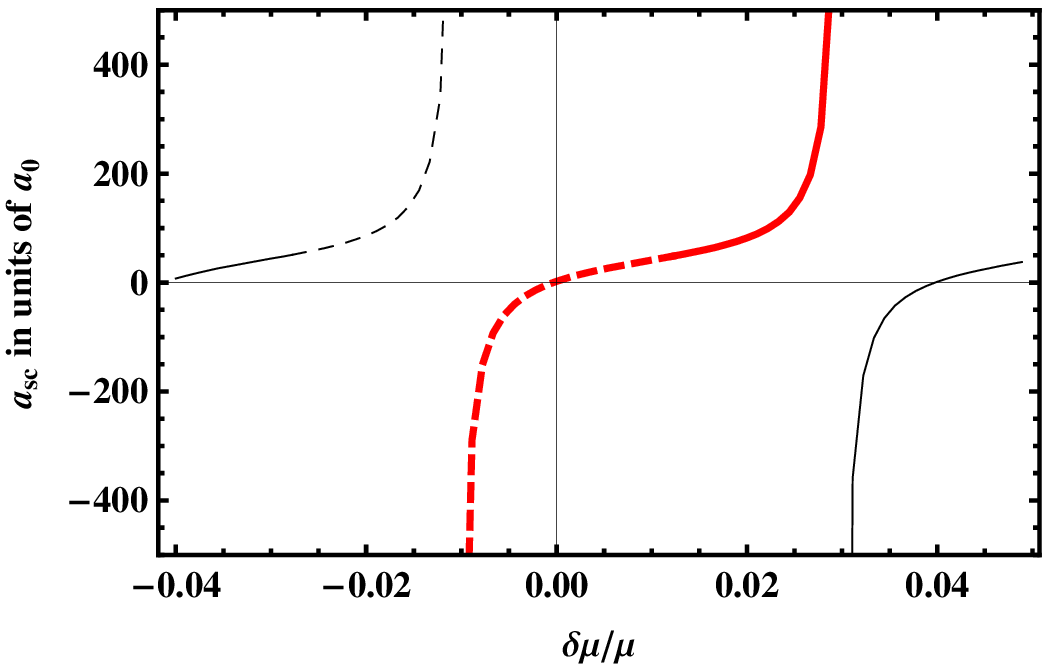}
     \end{minipage}
  {\bf (b)}\begin{minipage}[t]{0.45\textwidth}
     \vspace{0cm}\hspace{-1cm}
     \includegraphics[width=\textwidth]{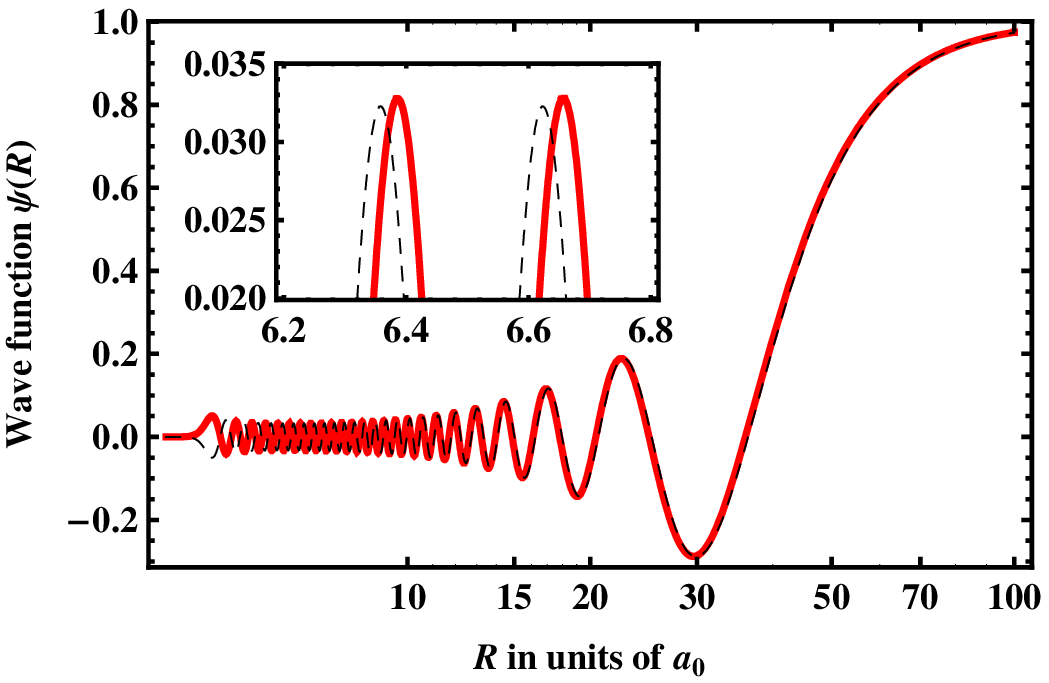}
     \end{minipage}
 \caption{{\footnotesize
     (Color online)
     (a) Scattering length $a_{\rm sc}$ as a function of the
     $\dm$-parameter of the mass variation. By constraining
     the variation to the thick (red) branch, any scattering length
     is reached while keeping the number of bound states $N_{\rm b}$ constant.
     By constraining it to the dashed (black/red) branch any scattering length
     is reached while $N_{\rm b}$ changes.
     (b) Resonant SC functions ($a_{\rm sc}=\infty$) at two different $\dm$-parameters
     $\dm\approx-0.01\mu$ (dashed), $\dm\approx 0.03\mu$ (thick, red).
     In order to make the relevant phase and amplitude difference at small 
     internuclear distances visible one of the wave function is multiplied 
     by $-1$ in the inset.
         }}
 \label{fig:aswf}
 \end{figure}

 As already mentioned, SC resonances are evoked by artificially shifting
 the least bound state or a virtual state across the threshold. 
 By turning a bound state into a virtual state or vice versa, the total number of 
 bound states $N_{\rm b}$ changes necessarily. This can be avoided by stopping
 the variation just before the bound or virtual states reach the threshold.
 Nevertheless it is possible to achieve any scattering length by moving between 
 two different SC resonances. This is illustrated by the example of the $\mu$
 variation in Fig.~\ref{fig:aswf}(a) where three resonant branches of the
 $a_{\rm sc}(\dm)$ curve are depicted. As discussed
 in~\cite{cold:gris07} the question arises, whether it is preferable to keep
 $N_{\rm b}$ constant or to change the variation parameter
 across a SC resonance as was done so far in this work.
%


\begin{figure}[!ht]
 \centering
  {\bf (a)}\begin{minipage}[t]{0.45\textwidth}
     \vspace{0cm}\hspace{-1cm}
     \includegraphics[width=\textwidth]{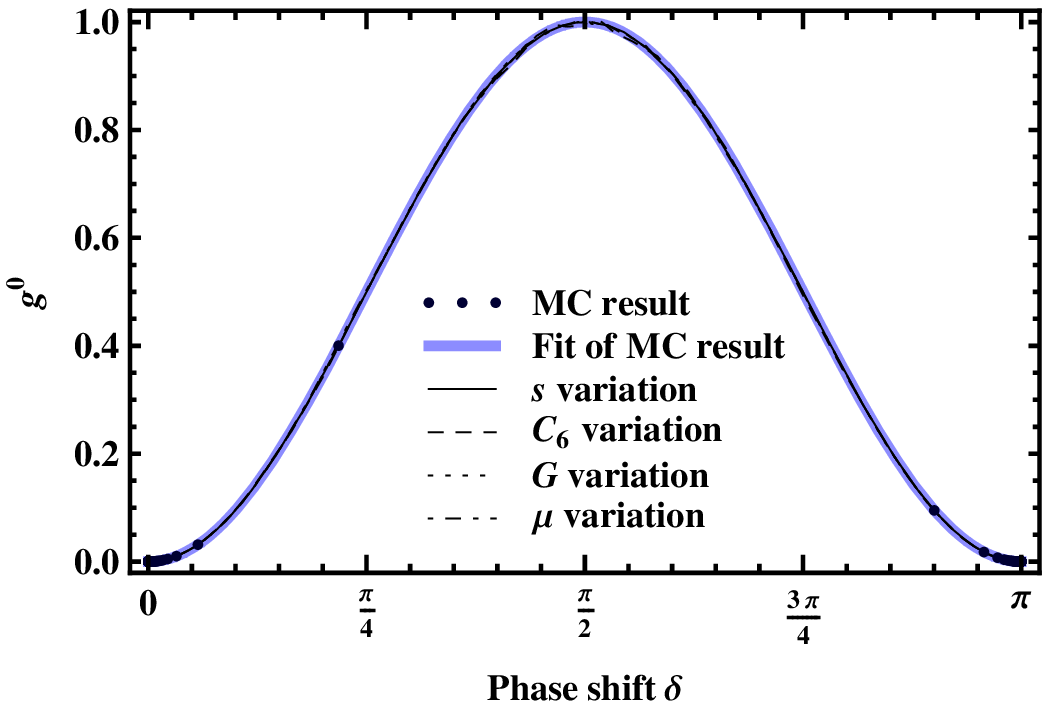}
     \end{minipage}
  {\bf (b)}\begin{minipage}[t]{0.45\textwidth}
     \vspace{0cm}\hspace{-1cm}
     \includegraphics[width=\textwidth]{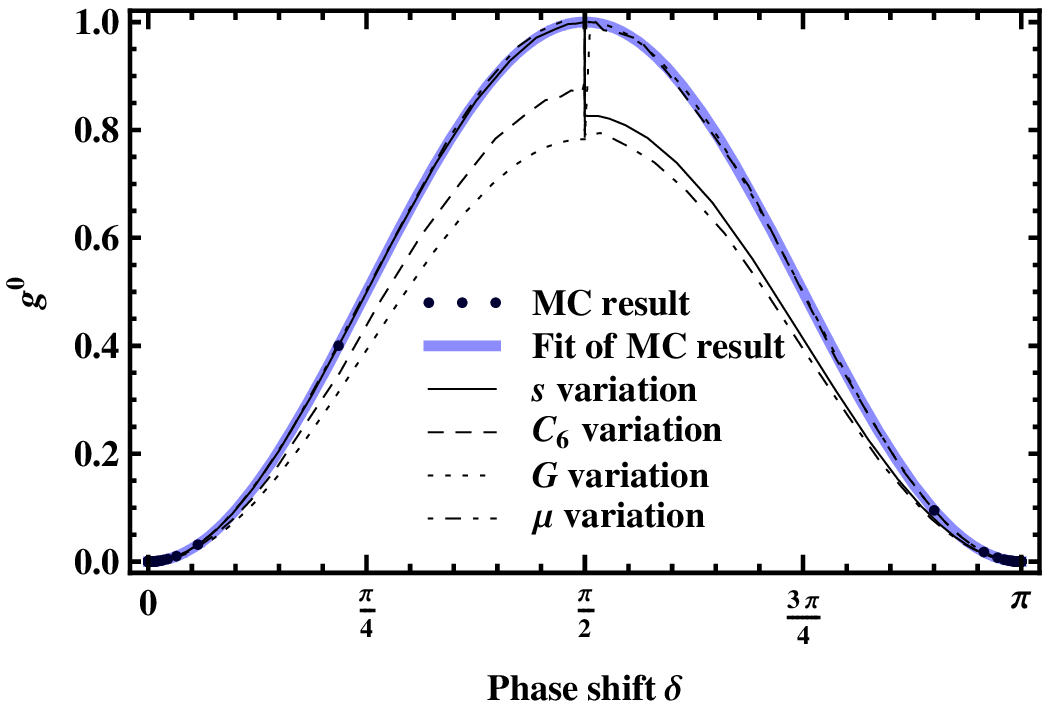}
     \end{minipage}
 \caption{{\footnotesize
     (Color online) Comparison of MC and SC results for the transition rate $I^0$ to the 
     absolute vibrational ground state relative to the respective 
     maximal transition rate as a function of the phase shift $\delta$.
     The different SC variation parameters are either varied around one SC
     resonance (a) or between two resonances staying on the same 
     $a_{\rm sc}(\dm)$ branch (b). The MC results are fitted, according  
     to the TC approximation (Eq.~(\ref{eq:IvSimp2Ch})). Note the $\sin^2$-like
     form of the functions.
         }}
 \label{fig:Rate_Branches}
 \end{figure}

 In Figs.~\ref{fig:Rate_Branches}(a) and (b) the relative transition rate is depicted as a
 function of the phase shift $\delta$. In comparison to a $1/a_{\rm sc}$-plot 
 (Fig.~\ref{fig:Comp_Rate_SC_MC}(b)), this allows an enlarged view on the 
 region of resonance where the phase $\delta$ suddenly crosses $\pi/2$. 
 In Fig.~\ref{fig:Rate_Branches}(a) the SC variations are performed in the same way as in
 Fig.~\ref{fig:Comp_Rate_SC_MC} around one SC resonance while changing
 $N_{\rm b}$. This results in a perfect agreement with the MC
 result (the deviations at small relative rates as shown in
 Fig.~\ref{fig:Comp_Rate_SC_MC}(a) are not visible on a linear scale of the
 relative rate). Furthermore, large values of the scattering length can be
 obtained by slight modifications of the SC Hamiltonian. By fixing
 $N_{\rm b}$, one has to stay on the same branch of the resonant curve  
 $a_{\rm sc}(\upsilon)$. This modifies the SC Hamiltonian strongly and can
 lead to a sudden change of the relative rate by some 30\% as is shown in
 Fig.~\ref{fig:Rate_Branches}(b).

 The reason for the sudden change of the wave function is twofold. In all
 cases the asymptotic behavior of the wave functions are the same at different
 resonant points corresponding the same $a_{\rm sc}$, but different
 Hamiltonians lead to a slightly different continuation of the wave function
 towards smaller distances. If the variation takes place at ranges larger than
 the equilibrium distance $R_e$ ($C_6,G$ and $\mu$ variation), the wave
 function around $R_e$ where it influences directly the transition rate can
 differ slightly in amplitude. Secondly, if the variation takes place around
 $R_e$ ($s$ and $\mu$ variation) the nodal structure of the SC wave functions
 differs for both resonant SC parameters, since the necessary phase shift is
 acquired in different ways. Both effects induce a ``step'' in the transition
 rate at $\delta=\pi/2$ as is visible in Fig.~\ref{fig:Rate_Branches}(b). The
 nodal shift can in principle change the transition rate more strongly than in
 the present case. Figure~\ref{fig:aswf}(b) compares the two wave functions of
 the $\mu$ variation at different resonant $\dm$ parameters. One can observe
 around $R\approx R_e$ both effects just described: the change of amplitude and
 the change of the nodal structure for different resonant variation
 parameters.

 To conclude, in order to calculate relative PA rates it
 should be in most cases preferable not to keep the number of bound states fixed and to
 avoid a sudden change of the SC scattering wave function while going over the
 resonance. The drawback is of course a sudden change of the wave function for
 small scattering lengths. But here the SC approaches show in any way
 differences to the MC result, such as a shift of the minimal transition rate
 along the $a_{\rm sc}$ axis (Fig.~\ref{fig:Comp_Rate_SC_MC}(a)). Noteworthy,
 for the energy spectrum analysis, as it was done, \eg, in~\cite{cold:gris09}
 for two atoms in an OL, it is more convenient to stay on the same SC resonant
 branch. The alternative variation with non-constant $N_{\rm b}$ does not
 influence the resulting energy spectrum. However, the disadvantage is that
 the numbering of the discrete levels should be changed across a SC resonance.

\section{Conclusion}
\label{sec:conclusion}

 We presented single-channel approaches that were able to reproduce both the
 long-range behavior of the open channel as well as the nodal structure and relative enhancement 
 of any singlet or triplet state of a multi-channel system in the presence of a
 magnetic Feshbach resonance. However, single-channel variations induce a shift of the nodal 
 structure not present in the multi-channel solution. Furthermore, the overall amplitude of the 
 wave function stemming from the asymptotical behavior can be slightly modulated by long and 
 intermediate-range variations.
 The $G$ variation, introduced in this work, showed to reproduce 
 the corresponding multi-channel components at short and long interatomic distances most accurately.

 As was demonstrated for the exemplary case of \LiRb scattering single-channel wave functions can 
 be used to study processes of molecule formation. 
 We examined the specific process of a direct one-photon photoassociation to the absolute vibrational 
 ground state of \LiRb and proved the applicability of the single-channel approaches to model this process.
 The effects of the nodal shift and the modulation of the amplitude lead to a discontinuity in the
 transition rate for either small scattering lengths, if varying the single-channel Hamiltonian over a
 resonance, or at large scattering lengths, if keeping the number of bound states constant. As was discussed, 
 a variation around a resonance of the single-channel Hamiltonian is preferable, since the point of
 minimal transition at small scattering lengths deviates in any way between multi-channel and 
 single-channel results. These deviations appear, however, at scattering lengths where
 the transition rate is negligible compared to the one at resonance.

 The general applicability of single-channel approaches was based on the two-channel approximation
 which reveals that the scaling of the open-channel wave function and the admixture of closed
 channels depends on the scattering length in a similar way.
 Additionally, by the help of this approximation one is able to reproduce exactly the
 multi-channel transition rate by adjusting two free parameters, that combine all details of the 
 transition process. 


 We can conclude that single-channel approaches are a suitable starting 
 ground to study molecular processes in regimes were full multi-channel calculations are too laborious.
 This is, \eg, the case, if the scattering takes place in an external trapping potential like
 an optical lattice, that in general couples relative and center-of-mass motions and 
 spoils the spherical symmetry. In most cases the trapping potential does not directly influence the 
 scattering wave function at short interatomic distances, but it induces an additional modulation of the 
 amplitude as a function of the scattering length. The examination of effects due to these modulations are 
 perfect candidates for the use of single-channel approaches.

 Since the nodal structure of either the singlet or the triplet components of the 
 multi-channel wave function is reproduced by single-channel approximations, also
 more complicated photoassociation schemes, exciting a range of 
 higher vibrational states, can be examined in the presence of a trapping potential.
 Furthermore, single-channel approaches allow to treat three and many-body collisions with reasonable
 numerical efforts in the presence of a magnetic Feshbach resonance.

 Of course, the presented single-channel approaches have also clear restrictions. For example, one has 
 to assume that the scattering energy and the background scattering length are sufficiently small.
 This condition can be spoiled for certain atomic systems and in deep external trapping potentials with
 significantly large ground state energy. Another problem can be caused by the energy dependence of 
 the scattering length especially for narrow Feshbach resonances.
 This energy dependence is not reflected by the current approaches. Furthermore, the multi-channel wave 
 function might behave differently compared to the single-channel one, if an energy variation is 
 induced by, \eg, ramping up an external trap. There exist single-channel approaches, which account for
 the energy dependence of the scattering length by a well-barrier pseudo-potential \cite{cold:DePa04}.
 However, like any pseudo potential it is unable to reflect the nodal structure of the scattering wave
 function at small internuclear distances.


 Recently Deiglmayr \etal observed for $^7$Li-$^{133}$Cs (at $B=0$) the exceptional case of a strong
 deviation of molecular channel functions from pure singlet or triplet behavior at small internuclear 
 distances \cite{cold:deig09}. Since spin-orbit coupling was neglected, they attributed this unusual 
 effect to strong hyperfine coupling but gave no reason, why this happens specifically for the 
 considered system.
 It is certainly interesting to further investigate this effect which would limit the 
 applicability of single-channel wave functions to predict, \eg, the relative transition rates to different 
 vibrational levels. 

 Apart from this unusual behavior it should be possible from the theoretical considerations presented 
 in this work to determine, whether and which single-channel approach is applicable for a specific system and molecular 
 process.

\section*{Acknowledgments}
%

 The authors are grateful to the {\it Stifterverband f\"ur die Deutsche
 Wissenschaft}, the {\it Fonds der Chemischen Industrie} and the 
 {\it Deutsche Forschungsgemeinschaft} (within {\it Sonderforschungsbereich 
 SFB\,450} and {\it Sa 936/2}) for financial support.


%
\end{document}